%% file: main.tex
\documentclass[twocolumn]{aastex62}

\usepackage[utf8]{inputenc}
\DeclareUnicodeCharacter{2212}{-}% support older LaTeX versions
\usepackage{graphicx}
\graphicspath{{./}}
\usepackage{amsmath}
\usepackage{afterpage}
\usepackage{enumitem}
\usepackage{xcolor}
\usepackage{float}
\setenumerate{itemsep=0mm}%noitemsep}
\usepackage{hyperref}

\usepackage{lineno}
\newcommand{\stamp}[0]{DELVE~2\xspace}
\newcommand{\Doublestamp}[0]{DELVE J0155$-$6815 (DELVE 2)\xspace}

\input{commands}

\shorttitle{DELVE Discovery of a New Ultra-Faint Stellar System}

\shortauthors{Cerny et al.}

\begin{document}

\reportnum{\footnotesize FERMILAB-PUB-20-485-AE}

\title{Discovery of an Ultra-Faint Stellar System  near the Magellanic Clouds with the DECam Local Volume Exploration (DELVE) Survey}

\input{authors}

\begin{abstract}
We report the discovery of a new ultra-faint stellar system found near the Magellanic Clouds in the DECam Local Volume Exploration (DELVE) Survey. This new system, \Doublestamp, is located at a heliocentric distance of $D_{\odot} = 71 \pm 4\kpc$, which places it at a 3D physical separation of 12 \kpc from the center of Small Magellanic Cloud (SMC) and 28 \kpc from the center of the Large Magellanic Cloud (LMC).
\stamp  is identified as a resolved overdensity of old ($\tau > 13.3\Gyr$) and metal-poor ($\feh = -2.0_{-0.5}^{+0.2}$  dex) stars with a projected half-light radius of $r_{1/2} = 21^{+4}_{-3} \pc$ and an absolute magnitude of $M_V = -2.1^{+0.4}_{-0.5}\magn$.
The size and luminosity of \stamp are consistent with both the population of recently discovered ultra-faint globular clusters and the smallest  ultra-faint dwarf galaxies.  However, its age and metallicity would place it among the oldest and most metal-poor globular clusters in the Magellanic system.
\stamp is detected in \Gaia DR2 with a clear proper motion signal, with multiple blue horizontal branch stars near the centroid of the system with proper motions consistent with the systemic mean. We measure the system proper motion to be $(\mu_{\alpha} \cos \delta, \mu_{\delta})= (1.02_{-0.25}^{+0.24}, -0.85_{-0.19}^{+0.18})$ mas~yr$^{-1}$. We compare the spatial position and proper motion of \stamp  with simulations of the accreted satellite population of the LMC and find that it is very likely to be associated with the LMC.
\end{abstract}


%-------------------------------------------------------------------------------

\section{Introduction}
\label{sec:intro}

The advent of large-scale digital sky surveys has revolutionized our understanding of the Milky Way and its satellite system. In particular, systematic searches of the Northern Hemisphere sky conducted with the Sloan Digital Sky Survey (SDSS; \citealt{York:2000}) first illuminated the Milky Way's lowest surface brightness populations, doubling the number of known dwarf galaxy satellites \citep[e.g.,][]{Willman2005AJ....129.2692W, Willman:2005,Belokurov2006ApJ...647L.111B,Belokurov2007ApJ...654..897B, Belokurov2009MNRAS.397.1748B, 2010ApJ...712L.103B,ZuckerApJ...650L..41Z,Zucker2006ApJ...643L.103Z}. 
The Pan-STARRS-1 \citep[PS1;][]{Chambers:2016} survey has further increased the coverage and depth of Northern Hemisphere surveys, resulting in the discovery of several new ultra-faint systems \citep[e.g.,][]{Laevens:2014,Laevens:2015a,Laevens:2015b}.
Furthermore, the advent of the Dark Energy Camera  \citep[DECam;][]{Flaugher:2015} on the 4m Blanco Telescope at the Cerro Tololo Inter-American Observatory (CTIO) in Chile has resulted in the discovery of a multitude of faint satellite galaxies and compact star clusters orbiting the Milky Way at surface brightnesses inaccessible to previous photographic surveys and SDSS. 
DECam searches covering $\roughly 5,000 \deg^{2}$ of the southern sky using data from the Dark Energy Survey \citep[DES;][]{Abbott:2005bi,DES:2016} resulted in the discovery of more than 20 new star cluster and dwarf galaxy satellites  \citep[e.g.,][]{Bechtol:2015,Koposov:2015cua,Kim:2015c,Drlica-Wagner:2015,Luque:2016,Luque:2018}. These efforts have continued through a number of recent community-led DECam surveys, including the Survey of the MAgellanic Stellar History \citep[SMASH; e.g.,][]{martin_2015_hydra_ii, Nidever2017AJ....154..199N}, the Magellanic SatelLites Survey \citep[MagLiteS; e.g.,][]{Drlica-Wagner:2016,Torrealba:2018a}, and the Magellanic Edges Survey \citep[MagES; e.g.,][]{Koposov:2018}, all of which have contributed to the census of Milky Way satellites, especially in the region of sky in the periphery of the Magellanic Clouds. 
In addition, several other surveys have found new ultra-faint dwarf galaxies including the Hyper Suprime-Cam Survey  \citep{Homma:2016, Homma:2017, Homma:2019}, VST ATLAS \citep{Torrealba:2016a, Torrealba:2016b}, and \Gaia \citep{Torrealba:2019}.

\par The detection of a large number of ultra-faint satellites ($\roughly 30$ in total) proximate to the Large and Small Magellanic Clouds (the LMC and SMC, respectively) has contributed to a growing body of theoretical and observational evidence suggesting that the LMC and SMC have brought their own satellite populations into the Milky Way \citep[e.g.,][]{DLB76,D'Onghia:2008a,Deason2015MNRAS.453.3568D,Sales2017MNRAS.465.1879S,DooleyPaper,Jethwa:2018, Kallivayalil:2018,Erkal:2019b,Jahn:2019,Nadler2020,Patel_2020}. In fact, the spatial distribution of the dwarf galaxy satellites discovered in the DES footprint alone excludes an isotropic spatial distribution for the Milky Way satellites at the $>3 \sigma$ level \citep{Drlica-Wagner:2015}.
Furthermore, high precision proper motion measurements from the \Gaia satellite \citep{Gaia:2018}, combined with radial velocity measurements, have allowed for the determination of these systems' 3D kinematics and orbital histories, linking  some dwarf galaxies and star clusters to the Magellanic system \citep[e.g.,][]{Kallivayalil:2018, Patel_2020}.
\par This developing picture of the Magellanic satellite system offers important insight into the Lambda Cold Dark Matter ($\Lambda \rm CDM$) paradigm, which predicts that galaxies form hierarchically across a wide range of mass scales. Furthermore, these low-mass, low-surface-brightness substructures can provide a wealth of information about their host halos --- in this case, the Magellanic Clouds. For example, these satellites have been used to place stringent constraints on the LMC/SMC mass \citep{EB2020_LMCmass}, and, through comparison with cosmological simulations, to trace the orbital history of the Clouds themselves \citep[e.g.,][]{Deason2015MNRAS.453.3568D, Jethwa:2016}. 
\par In this work, we present the discovery of an old, metal-poor, ultra-faint system, \Doublestamp, in the vicinity of the Magellanic Clouds. In addition to being a newly discovered member of the scarce population of old Magellanic stellar systems, this new system also occupies a region of size--magnitude space that makes it difficult to classify as either an ultra-faint cluster or dwarf galaxy. Due to this classification ambiguity, we refer to this new system as \stamp throughout this work, and consider several potential methods for elucidating the true nature of the system.
\par This paper is structured as follows.
In \secref{data}, we describe the DELVE survey observations and source catalogs used in this study. In \secref{search}, we detail our application of the \simple algorithm used to search for new ultra-faint systems in the periphery of the Magellanic Clouds, and present the detection of the candidate system, \Doublestamp.
In \secref{results}, we derive morphological and isochrone properties for \stamp, and present the detection of a clear proper motion signal for the system in data from \Gaia DR2. Lastly, in \secref{discussion}, we discuss the likely connection between \stamp and the Magellanic system, and consider how to classify \stamp as a stellar system.  
We conclude in \secref{conclusion}.

%-------------------------------------------------------------------------------

\section{Observations and Data}
\label{sec:data}

The DECam Local Volume Survey (DELVE; 2019A-0305) is a  multi-component, 126-night survey of the southern sky focused on studies of the satellite systems of the Milky Way, Magellanic Clouds, and several Magellanic-analog systems in the Local Volume. 
%Through three allocated years of observing, 
DELVE seeks to provide near-uniform, contiguous coverage of the southern sky with declination $\delta_{2000} < 0 \degree$ in the $g,r,i,z$ bands by combining all publicly available community DECam exposures with exposure times $>30$ seconds with $\roughly 20{,}000$ new exposures in regions of the sky not previously observed by DECam. 
DELVE is split into three observational components: DELVE-WIDE, a wide-area ($\roughly 10{,}500 \deg^{2}$) survey of the high-Galactic-latitude southern sky  to a depth of $g \sim 23.5\magn$; DELVE-MC: a contiguous survey of the Magellanic Cloud periphery ($\roughly 1{,}100 \deg^{2}$) to a depth of $g \sim 24.2\magn$; and DELVE-DEEP: a deep survey to a depth of $g \sim 25.0\magn$ around four isolated  Magellanic Cloud analogs in the Local Volume ($\roughly 150 \deg^{2}$), where the hierarchical prescriptions of $\Lambda\text{CDM}$ can be tested around intermediate--mass dark matter halos.
\par In \citet{MauCerny2020}, we presented results from an early satellite search over a $\roughly 4{,}000 \deg^{2}$ subregion from the WIDE survey component in the northern Galactic cap bounded by $b > 10\degree$ and $\dec < 0\degree$. In this work, we extend this search to a region of $ \roughly 2{,}200 \deg^2$  in the periphery of the Magellanic Clouds. Catalogs in this region were generated from community exposures and from new exposures from the DELVE WIDE and MC survey components.  
Our DELVE data set was constructed of $\roughly 9{,}000$ exposures in the periphery of the Magellanic Clouds and high-Galactic-latitude sky, including 1,000 new exposures from the first three semesters of DELVE observing (2019A, 2019B, and 2020A) and 8,000 community exposures publicly available before April 2020.\footnote{Community exposures were downloaded from the Science Archive hosted by NSF's National Optical-Infrared Astronomy Research Laboratory: \url{https://astroarchive.noao.edu/}.} 
Broadly, we began by selecting all available exposures with exposure times between 30 and 350 seconds in the region of sky south of the DES footprint ($\dec \lesssim -60 \degree$). In addition, we selected regions east and west of the DES footprint at Galactic latitude $10\degree < b < 20\degree$  with $\dec < -30 \degree$. 
We further excluded exposures in the densest central regions of the LMC and SMC, and removed exposures near the bright stars Canopus and $\beta$ Carinae.  
The primary contributors to the selected community exposures in the region are $\roughly 3,600$ exposures from MagLiteS \citep[2016A-0366, 	2018A-0242;][]{Drlica-Wagner:2016,Torrealba:2018a}, and $\roughly 550$ exposures from SMASH \citep[2013B-0440;][]{Nidever2017AJ....154..199N}, with other exposures sourced from more than 70 DECam observing programs. 
Approximately half of the selected exposures have exposure times of 90 seconds, consistent with the fact that DELVE-WIDE and MagLiteS performed $90$ second dithered exposures in the $g$, $r$, and $i$ bands. 
The remaining exposures, while initially selected based on a cut of 30 to 350 seconds, primarily include 267 and 333 second exposures from DELVE-MC and SMASH. 
\normalsize
\par We processed all exposures consistently with the DES Data Management (DESDM) pipeline \citep{Morganson:2018}. 
This pipeline achieves sub-percent-level photometric accuracy by performing full-exposure sky background subtraction \citep{Bernstein:2018} and calibrating based on custom, seasonally-averaged bias and flat images. 
The DESDM pipeline uses \verb|SourceExtractor| and \verb|PSFEx| \citep{Bertin:1996,Bertin:2011} on an exposure-level basis to automate source detection and photometric measurement. 
Stellar astrometry was calibrated against \Gaia DR2 \citep{Gaia:2018}, which provides 30 mas astrometric precision.
We calibrated the DELVE photometry by matching stars in each CCD to the ATLAS RefCat 2 catalogs \citep{Tonry:2018}, which  consists of measurements in the filter system used in PS1 DR1 \citep{Chambers:2016} and SkyMapper \citep{Wolf_2018}.  Photometric measurements from this catalog were transformed to the DECam $g, r, i, z$ filters before calibration using the following equations:
\begin{align*}
g_{\rm DECam} &= g_{\rm PS1} + 0.0994(g_{\rm PS1}-r_{\rm PS1}) - 0.0319 \\
r_{\rm DECam} &= r_{\rm PS1} -0.1335(g_{\rm PS1}-r_{\rm PS1}) + 0.0215\\
i_{\rm DECam} &= i_{\rm PS1} -0.3407(i_{\rm PS1}-z_{\rm PS1}) - 0.0013\\
z_{\rm DECam} &= r_{\rm PS1} -0.2575(r_{\rm PS1}-z_{\rm PS1}) - 0.0201,
\end{align*}\normalsize

The DELVE zeropoints have a root-mean-square scatter of $\roughly 0.01$ mag per CCD when compared to the DES zeropoints of these exposures \citep{Burke:2018}.
\par Next, we built a multi-band catalog of unique sources by matching source detections between the individual single-exposure catalogs following the procedure described in \citet{Drlica-Wagner:2015}. We cross-matched all sources detected in individual exposures using a $1''$ matching radius.
\par
\begin{figure}
\includegraphics[width=\columnwidth]{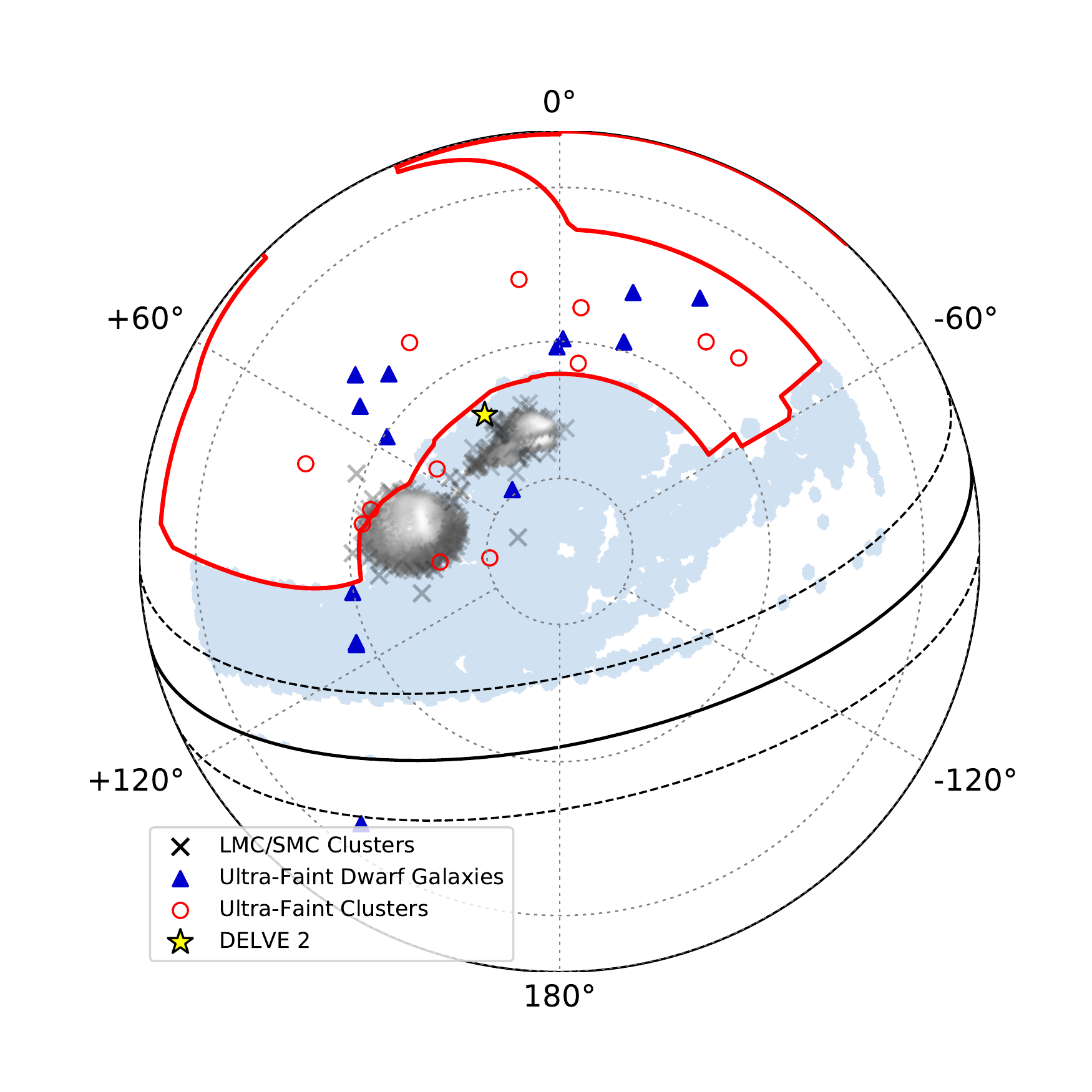}
\caption{Orthographic sky map of the Magellanic periphery region, including a selection of known satellites of the Milky Way and Magellanic Clouds. The blue shaded regions correspond to the $g$-band footprint of this search; i.e., these regions encompass the total area covered by searches in both the $g,r$ and $g,i$ bands. The red outline corresponds to the DES footprint. The dark black line corresponds to a Galactic latitude of $b = 0 \degree$, while the dashed lines correspond to $b = \pm 10 \degree$. LMC/SMC star clusters from the catalogs of \citet{Bica:2008} and \citet{Bica2020} are drawn in grayscale, with denser regions of clusters colored in whiter colors. The location of \stamp is indicated with a yellow star, lying just below the southern boundary of the DES footprint. 27 recently discovered ultra-faint star clusters and ultra-faint dwarf galaxies with possible Magellanic origins/associations, as listed in \citet{Bica2020}, are plotted as open red circles and filled blue triangles, respectively.}
\label{fig:skymap}
\end{figure}

 In total, the resulting catalog spanned an area $\roughly 2220 \deg^{2}$ with simultaneous coverage in $g,r$ and $\roughly 1880 \deg^{2}$ with simultaneous coverage in $g,i$. The difference between the $g,r$ coverage and the $g,i$ coverage is due to the availability of community data in the $r$ and $i$ bands. We include a map of the search region, in the context of known Magellanic satellite systems, in \figref{skymap}.
\par Lastly, we calculated extinction from Milky Way foreground dust for each object through a bilinear interpolation in $(\ra,\dec)$ to the maps of \citet{Schlegel:1998} with the rescaled normalization factor presented by \citet{Schlafly:2011}, assuming $R_V = 3.1$ and a set of coefficients $R_{\lambda} = A_{\lambda}/E(B-V)$ derived by DES for the $g$, $r$, and $i$ bands:  $R_g = 3.185$, $R_r = 2.140$, and $R_i = 1.571$ \citep{DR1:2018}.
Hereafter, all magnitudes quoted are corrected for interstellar  extinction.
% Throughout this work, we utilize the PSF magnitudes for each source measured in the exposure of the best image quality \FIXME{(highest S/N)} for each filter. 
We calculate the $10 \sigma$ limiting magnitude for the entire search region to be $g \sim 23.6 \text{ mag, } r \sim 23.3 \text{ mag, } \text{and } i \sim 22.6 \text{ mag}$.
%-------------------------------------------------------------------------------

\begin{figure*}
\center
% Diagnostic Plots for \stamp\\
\includegraphics[width=\textwidth]{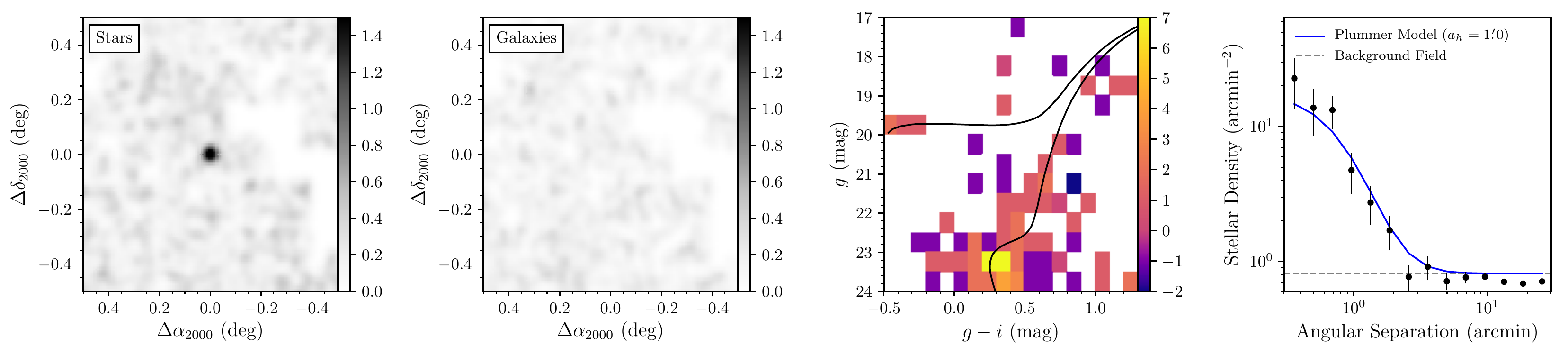}
\caption{Stellar density, galaxy density, Hess diagram, and radial density profile plots for \stamp. The visual inspection of the search results employed plots similar to these. 
Regions of missing coverage are visible as empty (white) regions in the density fields of the leftmost panels.
(Far Left) Isochrone-filtered stellar density field convolved with a Gaussian kernel of 1\arcmin.
(Middle Left) Isochrone-filtered background galaxy density field convolved with a Gaussian kernel of 1\arcmin.
(Middle Right) Color--magnitude Hess diagram corresponding to all foreground stars within 2.5 arcminutes of the centroid of \stamp minus all stars in a representative background region of the same area displaced to the east of the system (so as to avoid the region of incomplete coverage).
The best-fit \code{PARSEC} isochrone (derived in \secref{results}; \tabref{properties}) is shown in black.
White space indicates bins with no stars. The densest bin, colored in yellow, corresponds to an overdensity of main sequence turn-off (MSTO) stars. Multiple candidate blue horizontal branch stars are also visible.
(Far Right) Radial surface density profile of stars passing the isochrone filter; the errors are derived from the standard deviation of the number of stellar sources in a given annulus divided by the area of that annulus.
The blue radial profile curve corresponds to the best-fit Plummer model, assuming spherical symmetry, with $a_h=1\farcm0$ (see \secref{results}; \tabref{properties}).}
\label{fig:diagnostic}

\end{figure*}
\section{Satellite Search}
\label{sec:search}
\subsection{Methodology}
We search for new satellite candidates in the DELVE catalog described in  \secref{data} using the \simple\footnote{\url{https://github.com/DarkEnergySurvey/simple}} algorithm, which has been successfully applied for satellite searches on other DECam datasets \citep[e.g.,][]{Bechtol:2015,Drlica-Wagner:2015,Mau:2019,PaperI}, including most recently on early DELVE data in the northern Galactic cap, which resulted in the detection of the Centaurus I dwarf galaxy candidate and the DELVE 1 halo star cluster candidate \citep{MauCerny2020}. The \simple algorithm uses an isochrone matched-filter approach in color--magnitude space to enhance the contrast of spatial overdensities indicative of halo substructure against foreground Milky Way field stars. Using \simple, we performed two complementary searches, one of which utilized the $g,r$ bands, while the other utilized the $g,i$ bands.

We partitioned the catalog described above into \healpix \citep{Gorski:2005} pixels at $\nside=32$, corresponding to $\roughly 3.4\deg^2$ pixels. For each pixel, we selected stellar objects detected in both search filters using the morphological parameter $\var{spread\_model}$ and its associated error $\var{spreaderr\_model}$  \citep{2012ApJ...757...83D} by taking $|\var{spread\_model\_r}| < 0.003 + \var{spreaderr\_model\_r}$ for the $g,r$ band search and similarly $|\var{spread\_model\_i}| < 0.003 + \var{spreaderr\_model\_i}$ for the $g,i$ band search. In both cases, we also applied a magnitude selection of $g < 23.5\magn$ in order to reduce star-galaxy confusion and mitigate artificial density inhomogeneity in the data due to variations in survey depth. 

\par After star/galaxy separation, we applied a matched-filter template using a \code{PARSEC} isochrone \citep{Bressan:2012}, with metallicity $\metal=0.0001$ and age $\tau = 12 \Gyr$.
In each $\nside=32$ pixel and its eight neighboring pixels, we scanned our matched filter in distance modulus from $16.0 <\modulus < 23.5\magn$ in intervals of 0.5\magn, searching for spatial overdensities of old, metal-poor stars characteristic of halo star clusters and ultra-faint dwarf galaxies. At each distance modulus, we selected stars within 0.1\magn of the isochrone locus in color--magnitude space according to $\Delta (g-r) < \sqrt{0.1^2 + \sigma_g^2 + \sigma_r^2}$ (for the $g,r$ search) or $\Delta (g-i) < \sqrt{0.1^2 + \sigma_g^2 + \sigma_i^2}$ (for the $g,i$ search), where $\sigma_g$, $\sigma_r$, $\sigma_i$ are the photometric uncertainties for the $g,r,i$ band PSF magnitude measurements, respectively. We then smoothed the filtered stellar density field with a $2\arcmin$ Gaussian kernel, and identified local stellar density peaks by iteratively increasing a density threshold until fewer than ten disconnected peaks were detected. Lastly, for each of the identified density peaks, we computed the Poisson significance of the observed number of filtered stars relative to an annular background field and compiled a candidate list.
%%%%%%%%%%%%%%%%%%%%%%%%%%%%%%%%%%
\begin{figure*}
\center

\includegraphics[width=\textwidth]{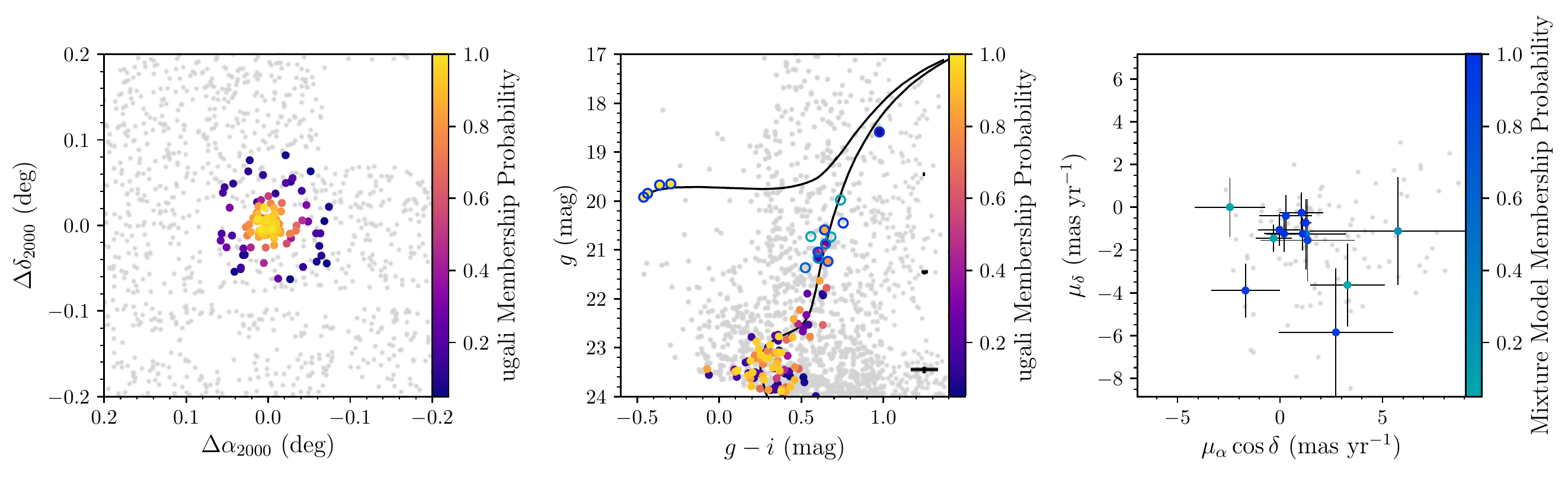}
\caption{
    Spatial distribution map, color--magnitude diagram, and proper motion plot for \stamp, with the former two colored by \ugali membership probability ($p_{\rm \ugali}$) and the latter colored by mixture model membership probability ($p_{\rm MM}$).
   (Left) Spatial distribution of stars with $g < 24.0\magn$ and $i < 23.5\magn$  in a $0.16\deg^2$ area region around the centroid of \stamp; possible member stars, defined as those with $p_{\rm \ugali} > 0.05$, are colored by their \ugali membership probability. Stars with $p_{\rm \ugali} \leq 0.05$ are shown in gray. 
   (Center)  Color--magnitude diagram of the stars shown in the left panel, applying the same magnitude cuts. We include representative photometric error bars sampled at three different $g$-band magnitudes (19.5 mag, 21.5 mag, 23.5 mag) in black. These error bars are positioned at an arbitrary location along the color axis.
    The best-fit \code{PARSEC} isochrone is drawn in black.
    Four blue horizontal branch stars are identified as highly probable members of \stamp and are clustered very closely to the centroid of the system.
    Stars cross-matched with \Gaia DR2 with $p_{\rm MM} > 0.05$ are outlined by their mixture model membership probability.
    (Right) \Gaia proper motions for stars cross-matched with the DELVE discovery data.
    Stars with $p_{\rm MM} > 0.05$ are colored by their mixture model membership probability, and stars with $p_{\rm MM} \leq 0.05$ that pass an isochrone filter are shown in gray.
}
\label{fig:membership}
\end{figure*}
\subsection{Detection of \stamp}
We visually inspected diagnostic plots for all candidates identified at Poisson significance $ \geq 5.5\sigma$ in at least one of the two searches and all candidates simultaneously identified at the $\geq 5 \sigma$ level in both searches. In addition to recovering all known dwarf galaxies within the search footprint, including Carina \citep{Cannon:1977}, Carina~II and Carina~III \citep{Torrealba:2018a},   Hydrus~I \citep{Koposov:2018}, and Pictor~II \citep{Drlica-Wagner:2016}, and a number of known LMC/SMC periphery clusters, one previously unknown stellar overdensity was identified in the constellation Hydrus. This candidate, \Doublestamp, was identified as a compact density peak at a Poisson significance of $9.1 \sigma$ in the $g,r$ band search and $5.65\sigma$ in the $g,i$ band search. 
\par In \figref{diagnostic}, we present diagnostic plots for \stamp similar to those inspected in the \simple search results. As is visible in the two lefthand panels, \stamp was detected as a clear overdensity against the field of foreground stars (far left) and background galaxies (middle left); however, \stamp was discovered in a region of our catalog with only a single tiling in each of $g,r,i$ bands, leading to a conspicuous lack of coverage to the west of the system. Our diagnostic Hess diagram (middle right panel) indicated a clear overdensity of main sequence turn off stars at the position of \stamp, and featured several potential blue horizontal branch stars. Lastly, the far right panel of \figref{diagnostic} shows the radial distribution of isochrone-filtered stars with respect to the centroid of \stamp.

\par Given the large number of previously discovered stellar systems associated with the Magellanic Clouds found in the region of sky near \stamp, we verified the novelty of this discovery through cross-matching the coordinates of the \stamp system with multiple independently maintained astronomical datasets/databases.
Adopting the discovery coordinate centroid of $(\ra, \dec) = (28.77\degree,-68.27\degree)$, we found no known stellar systems within 10 arcminutes in SIMBAD \citep{SIMBAD}, the catalog of known confirmed or candidate Milky Way satellite dwarf galaxies presented in Table 2 of \citet{PaperI}, the catalog of Magellanic Cloud star clusters and stellar associations presented by \citet{Bica:2008} and its comprehensive recent update of SMC/Magellanic Bridge stellar systems \citet{Bica2020}, which includes a list of 27 ultra-faint Magellanic-region systems. While the catalog of LMC clusters presented in \citet{Bica:2008} has not yet been updated to reflect more recent cluster discoveries, we note that \stamp's position (depicted visually in \figref{skymap}) places it outside the search regions for many recent efforts dedicated to searching for outer LMC satellites, including those conducted by DES \citep{Pieres2016}, the Optical Gravitational Lensing Experiment (OGLE\footnote{Candidate clusters identified by OGLE in the SMC periphery and Magellanic bridge by \citet{2017AcA....67..363S} are included in the catalog of \citet{Bica2020}.}; \citealt{OGLE_LMC}), and the YMCA (Yes, Magellanic Clouds Again) and STEP (The SMC in Time: Evolution of a Prototype interacting late-type dwarf galaxy) projects conducted on the VLT Survey Telescope \citep{2020arXiv200700341G}.
\par We proceed with $g,i$ band tiling for further analysis of this candidate, as the effective exposure time and stellar PSFs for the $i$-band exposure at this location are better than the $r$-band exposure, 
which primarily resulted in more reliable star/galaxy separation.
Since the $r$ and $i$ exposures are positioned identically on-sky, there is no difference in coverage between the two bands. 
While the detection significance was higher for the $g,r$ search, we found that the calculated test statistic (TS) from our parameter fit (\secref{results}) was higher when using the $g,i$ bands. We note that the initial difference in detection significance was due to a relatively poor determination of the distance modulus in the $g,i$ band search. Furthermore, we found that the best-fit parameters derived using the $g,r$ band data were consistent within the stated uncertainties of the $g,i$ band results presented in \secref{results}.
%-------------------------------------------------------------------------------

\section{Properties of \stamp}
\label{sec:results}

In the following subsections, we characterize the morphology, stellar populations, distance, and proper motion of \stamp.
We present the most probable values of these parameters with their associated uncertainties in \tabref{properties}.

\subsection{Morphological and Isochrone Parameters}
To fit the morphological and isochrone parameters of \stamp, we utilized the Ultra-faint Galaxy Likelihood (\ugali) software toolkit\footnote{\url{https://github.com/DarkEnergySurvey/ugali}} \citep{Bechtol:2015,Drlica-Wagner:2015}, which uses an unbinned Poisson maximum-likelihood formalism to derive best-fit parameters and identify probable member stars for resolved stellar  systems.\footnote{Appendix~C of \citet{PaperI} describes the statistical formalism implemented by \ugali.} We modeled the spatial distribution of stars with a \citet{Plummer:1911} profile, accounting for the incomplete coverage near the system. A template \citet{Bressan:2012} isochrone was fit to the observed color--magnitude diagram.  We simultaneously fit the centroid right ascension and declination (\ra and \dec, respectively), extension along the semi-major axis ($a_h$), ellipticity (\ellip), and position angle (P.A.) of the Plummer profile, and the  distance modulus  (\modulus), age (\age), and metallicity (\metal) of the isochrone, and derived the posterior probability distributions for each parameter using the affine-invariant Markov Chain Monte Carlo (MCMC) ensemble sampler, \code{emcee} \citep{Foreman-Mackey:2013}. From these properties, we then derived estimates of the Galactic longitude and latitude ($\ell$ and $b$, respectively), the azimuthally averaged angular and physical half-light radii ($r_h$ and $r_{1/2}$, respectively), the average surface brightness within the half-light radius ($\mu$), the heliocentric distance ($D_{\odot}$), the 3D Galactocentric distance ($D_{\rm GC})$ between \stamp and the Galactic center (assumed to be at $D = 8.178\kpc$; \citealt{Abuter:2019}), the total stellar mass integrated along the best-fit isochrone ($M_{*}$), absolute integrated visual magnitude ($M_{V}$), and the metallicity (\feh).  The \ugali membership probability ($p_{\rm \ugali}$) of each star comes from the Poisson probability that a star belongs to \stamp given its location relative to the best-fit spatial model, proximity to the best-fit isochrone in color-magnitude space, photometric measurement uncertainty, and the local imaging depth. We calculated the sum of the \ugali membership probabilities across all stars, $\sum_{i} p_{i, {\rm \ugali}} = 59$, indicating a significant number of likely member stars with $g < 23.5\magn$.
\par In  \figref{membership}, we plot the spatial distribution, color-magnitude diagram, and proper motion vector-point diagram (see \secref{proper_motion}) for \stamp.  In the left two panels, all stars with $p_{\rm \ugali} > 0.05$ are colored by their \ugali membership probability. In the middle panel, the derived best-fit \code{PARSEC} isochrone is drawn in black, and we circle stars identified as likely system members based on \Gaia DR2 proper motions (shown in the righthand panel; see  \secref{proper_motion}).

\begin{deluxetable}{l c c}[th!]
\tablecolumns{3}
\tablewidth{0pt}
\tabletypesize{\footnotesize}
\tablecaption{\label{tab:properties} Morphological, isochrone, and proper motion parameters for \stamp based on the $g,i$ band data.
}
\tablehead{
\colhead{Parameter} & \colhead{Value} & \colhead{Units}}
\startdata
\ra  & $28.772^{+0.006}_{-0.005}$& deg \\
\dec & $-68.253^{+0.002}_{-0.002}$ & deg\\
$\ell$  & 294.236 & deg \\
$b$  & -47.789 & deg \\
$a_\text{h}$ & $1.04^{+0.19}_{-0.15}$ & arcmin \\
$r_\text{h}$  & $1.02^{+0.18}_{-0.15} $ & arcmin  \\
$r_{1/2}$  & $21^{+4}_{-3}$ & pc  \\
\ellip & $0.03^{+0.15}_{-0.03}$  & ... \\
\PA  & $74^{+84}_{-40}$ &  deg \\
\modulus  &  $19.26^{+0.03}_{-0.03}\pm 0.1$\tablenotemark{a} & mag\\
$D_{\odot}$  & $71\pm4$ & kpc\\
\age  & $>13.3 $\tablenotemark{b} & Gyr \\
\metal & $0.00015^{-0.0001}_{+0.0001}$  & ... \\
$\sum_i p_{i, {\rm \ugali}}$ & $59^{+18}_{-10}$ & ... \\
\TS & 181 & ... \\[-0.5em]
\multicolumn{3}{c}{\hrulefill} \\
$M_V$ & $-2.1^{+0.4}_{-0.5}$ \tablenotemark{c} & mag  \\
$M_{*}$ & $880^{+120}_{-150} $\tablenotemark{d} & ${\rm M}_{\odot}$ \\
$\mu$  & 28.2 & mag~arcsec$^{-2}$\\
\feh  & $-2.0_{-0.5}^{+0.2}$ \tablenotemark{e}& dex \\
$E(B-V)$ & 0.024 & mag \\
$D_{\rm GC}$  & $69\pm4$ & kpc  \\[-0.5em]
\multicolumn{3}{c}{\hrulefill} \\
$\mu_{\alpha} \cos \delta$  & $1.02_{-0.25}^{+0.24}$ & mas~yr$^{-1}$  \\
$\mu_{\delta}$ & $-0.85_{-0.19}^{+0.18}$ & mas~yr$^{-1}$ \\
$\sum_i p_{i, {\rm MM}}$ & $9.5_{-0.3}^{+1.1}$ & ...\\[+0.5em]
\enddata
\tablecomments{Uncertainties were derived from the highest density interval containing the peak and 68\% of the marginalized posterior distribution.}
\tablenotetext{a}{We assume a systematic uncertainty of $\pm0.1$ associated with isochrone modeling \citep{Drlica-Wagner:2015}.}
\tablenotetext{b}{The age posterior peaks at the upper bound of the allowed parameter range ($13.5\Gyr$); thus, we quote a lower limit at the 84\% confidence level.}
\tablenotetext{c}{The uncertainty in $M_V$ was calculated following \citet{Martin:2008} and does not include uncertainty in the distance.}
\tablenotetext{d}{We note that our estimate of $M_{*}$ does not account for a mass contribution from possible blue straggler stars due to the difficulty of distinguishing them from the SMC foreground.}
\tablenotetext{e}{Our estimate of \feh is derived from the best-fit \code{PARSEC} isochrone following the procedure described in Section 3 of \citet{Bressan:2012} assuming a solar metallicity of $\metal_\odot = 0.0152$.}

\vspace{-3em}
\end{deluxetable}
\subsection{Proper Motion}
\label{sec:proper_motion}

To verify the detection of \stamp and measure the proper motion, we cross-matched stars within 0\fdg5 of the system centroid with the \Gaia DR2 catalog \citep{Gaia:2018}.
% To see if stars in each system show coherent systemic motion on the sky, we cross-matched stars in the DELVE-WIDE catalog to the \Gaia DR2 catalog \citep{Gaia:2018} to measure their proper motions.
The stellar sample was filtered by selecting stars consistent with zero parallax ($\varpi - 4 \sigma_\varpi \leq 0$), small proper motions (i.e., removing stars that would be unbound to the Milky Way if they were at the distance of \stamp), and a color--magnitude selection of 0.1\magn in $g$--$i$  from a best-fit isochrone with metallicity ${\rm [Fe/H]}=-2.2$ and age $\age=13.5\Gyr$.
% We note that \Gaia DR2 has a limiting magnitude of $G \sim 21\magn$ \citep{Gaia:2018}, which is significantly shallower than that of the DELVE-WIDE dataset.

To determine the proper motion of the satellite, we applied a Gaussian mixture model  \citep{Pace:2019}.
Briefly, the mixture model separates the likelihoods of the satellite and the foreground stars, decomposing each into a product of spatial and proper motion likelihoods.
%Note that the foreground here is composed of the Milky Way, Large Magellanic and Small Magellanic Clouds.
Stars that are closer to the centroid of the satellite are given higher weight based on the best-fit stellar distribution and stars well outside the satellite help determine the Milky Way foreground proper motion distribution.
In contrast to \citet{Pace:2019}, we varied $a_\text{h}$ and assume a Gaussian prior based on the \ugali results ($a_\text{h}=1.02\pm0.17 \, {\rm arcmin}$).  In addition we utilized two components in the foreground population to model the LMC/SMC and Milky Way populations.
The MultiNest algorithm \citep{Feroz2008MNRAS.384..449F, Feroz2009MNRAS.398.1601F} was used to determine the best-fit parameters, including the proper motions of the satellite and of the Milky Way foreground stars.
The mixture model membership probability ($p_{\rm MM}$) of each star was calculated by taking the ratio of the satellite likelihood to the total likelihood from the posterior distribution (see \citealt{Pace:2019} for more details).

% We derive a proper motion of $(\mu_{\alpha} \cos \delta, \mu_{\delta}) = (0.96^{+0.21}_{-0.22}, -0.91^{+0.17}_{-0.17}) \mas~\yr^{-1}$  and $(\mu_{\alpha} \cos \delta, \mu_{\delta}) = (1.05^{+0.24}_{-0.23}, -0.85^{+0.17}_{-0.17}) \mas~\yr^{-1}$ for  $g$--$r$ and $g$--$i$ color-magnitude selections, respectively.
We derive a systemic proper motion of  $(\mu_{\alpha} \cos \delta, \mu_{\delta}) = (1.02^{+0.24}_{-0.25},\allowbreak -0.85^{+0.18}_{-0.19}) \mas~\yr^{-1}$.
In the rightmost panel of \figref{membership}, we color candidate members stars with $p_{\rm MM} > 0.05$ by their mixture model membership probability, and foreground stars with $p_{\rm MM} \leq 0.05$ are shown in gray.
Stars cross-matched between the DELVE discovery and \Gaia DR2 with $p_{\rm MM} > 0.05$ are outlined in the color-magnitude diagram in the central panel of \figref{membership}.
We define the sum of the mixture model membership probabilities as $\sum_{i} p_{i, {\rm MM}}$, and find $\sum_i p_{i, {\rm MM}} = 9.5_{-0.3}^{+1.1}$.
If we assume that \stamp has a \citet{Chabrier:2001} initial mass function with an age of 13.5 Gyr and [Fe/H] of -2.2, then we predict that we should observe $N=4_{-2}^{+3}$ stars brighter than $g\sim21\magn$ in \Gaia based on 1000 \ugali simulations.  
The difference between the (lower) predicted number of stars and the observed number is driven by the fact that the simulations predict a lower number on the horizontal branch than the four stars we observe.

\subsection{Checking for Known RR Lyrae Variable Stars}
\par In an attempt to further constrain the distance to  \stamp, we searched the OGLE \citep{Soszynski2019AcA....69...87S} and \Gaia \citep{Clementini2019A&A...622A..60C}  catalogs for potential known RR Lyrae (RRL) variable stars associated with the system. These stars obey a well-constrained period-luminosity-metallicity relation, making them excellent standard candles for tracing the distances to old stellar populations, where they are often found.  We found that there are 7 and 3 RRL stars within 1\degree of \stamp in the OGLE and \Gaia catalogs, respectively.  However, the closest RRL variable to \stamp is $\sim21$ arcminutes away in projection (far beyond the maximum observed extent of the system) and all RRL stars are at closer heliocentric distances. Therefore, we conclude that these nearby RRL stars are consistent with either the foreground Milky Way stellar population or stars in the outskirts of the SMC and are not likely to be associated with \stamp. However, we note that all known Milky Way dwarf galaxy satellites fainter than $M_{V} = -3.0$ have one or fewer known RRL \citep{Martinez-Vazquez:2019}, and thus the lack of known RRLs at the position of \stamp is not particularly surprising, provided the system is a dwarf galaxy.

\section{Discussion}
\label{sec:discussion}
In the previous sections, we have presented the discovery of \stamp and have characterized its morphological properties, stellar population, and systemic proper motion. In the following sections, we utilize this information to discuss \stamp's potential association with the Magellanic Clouds and its classification as a stellar system.

\begin{figure}

\center
\includegraphics[width=\columnwidth]{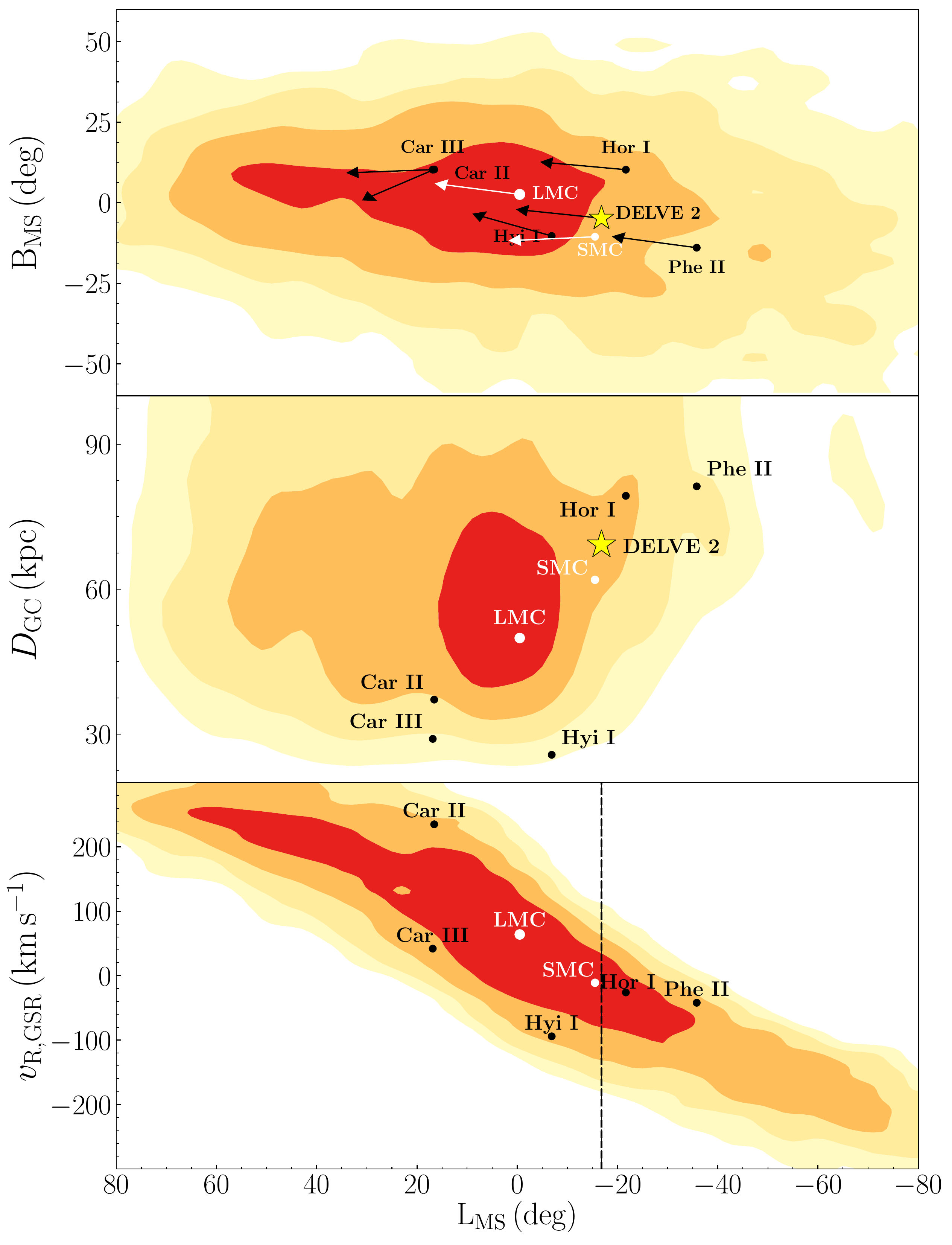}

\caption{
    Smoothed relative density of simulated LMC satellites from \citet{Jethwa:2016}, normalized to unity in each plot, displayed in Magellanic Stream coordinates.
    \stamp is shown as a yellow star along with five likely LMC satellite galaxies \citep[Hor~I, Car~II, Car~III, Hyi~I, Phe~II;][]{Kallivayalil:2018,Erkal:2019b}, shown as black circles. The LMC and SMC are shown as white circles. Note that Car~II and Car~III are spatially coincident in projection but have different proper motion vectors, Galactocentric distances, and velocities.
    (Top) Proper Motion of the LMC, SMC, and the five satellites. Arrows indicate the solar-reflex-corrected proper motions of each system (no physical meaning is attributed to the magnitudes of these arrows).
    (Middle) Galactocentric distances ($D_{\rm GC}$)  of the five likely LMC satellites and \stamp.
    (Bottom) Line-of-sight velocities in the Galactic standard of rest ($v_{\rm R,GSR}$) for 5 of the Magellanic dwarf galaxy satellites, in addition to the LMC and SMC. The black dashed line represents the MS longitude of \stamp.}
\label{fig:lmc}
\end{figure}

\subsection{Association with the Magellanic Clouds}

The position of \stamp in projection relative to the SMC and LMC ($\roughly 6.9 \degree$ and $\roughly 18.2 \degree$ respectively) raises the immediate question of whether a physical association exists between the systems.
We calculate the 3D separation between the centroid of \stamp and the SMC, $D_{\rm SMC}$, to be $\roughly 12.1 \kpc$ assuming ($\alpha_{\rm SMC}, \delta_{\rm SMC}) =  (13.187\degree,-72.829\degree)$  in celestial coordinates and a SMC distance of 61.94 kpc \citep{dGB_smc}. Similarly, we calculate the 3D separation between \stamp and the LMC to be $\roughly 28.30$ kpc, assuming an LMC centroid of ($\alpha_{\rm LMC}, \delta_{\rm LMC}) = (80.90\degree,-68.74\degree)$ \citep{Skymapper:2020} and an LMC distance of 49.89 kpc \citep{Grijs_Distance}. This places the system beyond recent estimates for the tidal radius of the SMC \citep[e.g., 5 kpc; ][]{2020SMC}, implying the gravitational influence of the significantly more massive LMC is likely to be stronger at the position and distance of \stamp. Nonetheless, the system also resides beyond many recent estimates of the tidal radius of the LMC --- for example, $ 22.3 \pm 5.2 \text{ kpc}$ from \citet{van_der_Marel_2014}. These calculated separations, while fairly large compared to known cluster satellites, preliminarily suggest an association with the Magellanic Cloud satellite system, which we explore further by considering the proper motion signal detected in \Gaia DR2. 
\par To probe the relationship between \stamp and the Magellanic Clouds, we compare the proper motion derived in Section~\ref{sec:proper_motion} to the LMC infall models of \citet{Jethwa:2016} in \figref{lmc}. 
We plot the spatial position in Magellanic Stream coordinates \citep{Nidever:2008} of \stamp, the LMC, SMC, and five known ultra-faint galaxies suggested to be associated with the Magellanic system \citep{Kallivayalil:2018,Erkal:2019b, Patel_2020}  over the numerically simulated LMC tidal debris, and visually highlight the direction of their solar-reflex-corrected proper motion vectors and that of \stamp. 
These five ultra-faint galaxies are Horologium~I, Carina~II, Carina~III, Hydrus~I, and Phoenix~II, with proper motion measurements coming from \citet{Kallivayalil:2018} and \citet{Pace:2019}.
While the classical dwarf galaxies Carina and Fornax have also been suggested to be LMC satellites \citep[e.g.,][]{Pardy:2019}, we do not include them since more detailed orbit modelling by \citet{Erkal:2019b} and \citet{Patel_2020} found that neither system is likely to be an LMC satellite. 
\par As is visually apparent in the top panel of \figref{lmc}, the proper motion of \stamp is consistent with those of the LMC and the SMC.
\stamp is trailing the Magellanic system, similar to the ultra-faint dwarfs Hor~I and Phe~II (as measured by \citealt{Kallivayalil:2018}). \stamp lies in a region with reasonably high simulated LMC satellite density, with a Galactocentric distance between that of the SMC and Hor~I. Based on the simulation data, we find that \stamp is most likely to be associated with the Magellanic system if its line-of-sight velocity in the Galactic standard of rest is within the range $-80 \text{ km s}^{-1} \lesssim v_{\rm R, GSR} \lesssim 50 \text{ km s}^{-1}$. While a measurement of the line-of-sight velocity is required to confirm membership in the Magellanic system, we find it to be highly likely that \stamp is a member based on the available data in comparison to simulations and known satellites.
\par As an additional check, we integrated the orbit of \stamp backwards in time to determine whether it was originally an LMC satellite as in \cite{Erkal:2019b}.
In particular, we Monte Carlo sampled the present-day proper motions and distance 10,000 times from the values in this work. For each realization, we uniformly sampled the radial velocity between -500 to 500 $\text{ km s}^{-1}$ and sampled the LMC's radial velocity, distance, and proper motions from their observed values \citep{lmcrv, lmcpm, Pietrzynski:2013}. 
The LMC was modelled as a Hernquist profile \citep{Hernquist:1990} with a mass of $1.5\times10^{11} M_\odot$ and a scale radius of 17.13 kpc \citep[consistent with the results of][]{Erkal:2019a}. The Milky Way was modelled with a potential nearly identical to \texttt{MWPotential2014} from \cite{galpy}; the only difference was that the bulge was replaced with a Hernquist profile with a mass of $5\times10^9 M_\odot$ and a scale radius of $0.5$ kpc.
\par \stamp was integrated backwards in the combined presence of the Milky Way and LMC for 5 Gyr, significantly before the LMC's accretion onto the Milky Way. At the end of the integration, we determined whether \stamp was originally bound to the LMC. Given the 10,000 iterations, we then estimated the probability that it was bound as a function of radial velocity.
We find a large range of radial velocities ($-150$ $\text{ km s}^{-1} < v_{\rm R,GSR} < 80$ $\text{ km s}^{-1}$ for which \stamp has a $>50\%$ chance of being an LMC satellite. 
The peak probability of $0.81$ occurred at $v_{\rm R,GSR} \sim -5\text{ km s}^{-1}$. 
This result is broadly consistent with the predictions from the forward-modelled population of LMC satellites discussed above \citep[i.e., from][]{Jethwa:2016}.
Thus, \stamp is a promising LMC satellite candidate and future radial velocity measurements will likely be able to confirm its membership.

\begin{figure*}
\center
\includegraphics[width=\textwidth]{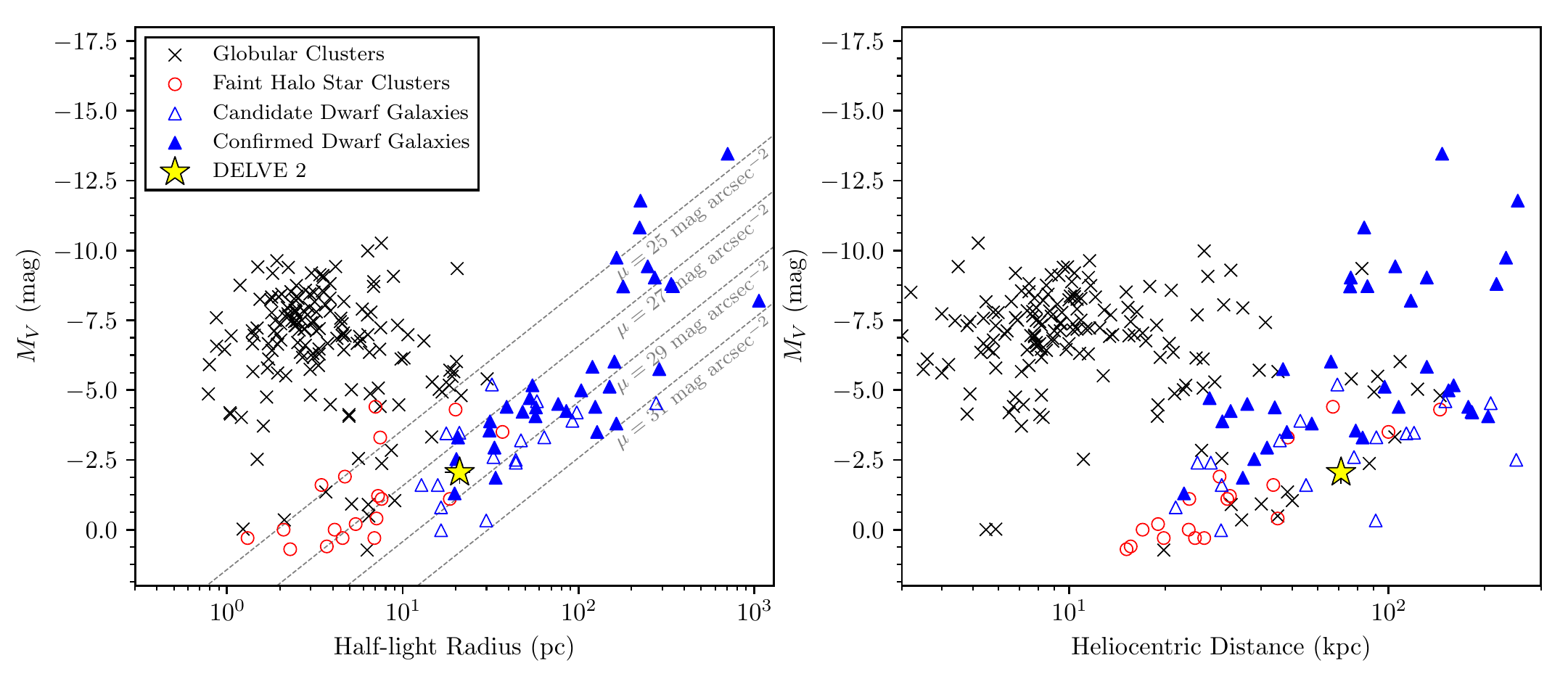}
\caption{
    (Left) Absolute magnitude vs. azimuthally averaged physical half-light radius for dwarf galaxy satellites of the Milky Way/LMC/SMC system, \citep[unfilled and filled blue triangles for candidate and confirmed dwarf galaxies, respectively;][and references therein]{PaperI}, globular clusters \citep[black crosses;][]{Harris:1996}, and recently discovered halo star clusters with $D_{\odot} > 10$ kpc. \citep[unfilled red circles;][]{Fadely:2011,Munoz:2012,Balbinot:2013,Belokurov:2014,Laevens:2014,Kim:2015a,Laevens:2015b,Kim:2016,Luque:2016,Luque:2017,Luque:2018,Koposov:2017,Mau:2019,Torrealba:2019, MauCerny2020}.
    \stamp is displayed as a yellow star.
    Lines of constant surface brightness are drawn as diagonal dashed grey lines.
    (Right) Absolute magnitude vs.\ heliocentric distance of ultra-faint stellar systems in the Milky Way/LMC/SMC system.
    \stamp occupies the ambiguous regime between recently discovered Milky Way halo star clusters and dwarf galaxies in this three-dimensional parameter space.
}
\label{fig:satellites}
\end{figure*}

\subsection{Tidal Disruption from the Magellanic Clouds?}
Given the relative proximity of \stamp to the Magellanic Clouds it is reasonable to explore whether the system might be undergoing tidal disruption due to the LMC, even if no obvious tidal features are visible in the spatial distribution of stars presented in \figref{membership}. 
To probe the survivability of \stamp against the strong gravitational forces of the Magellanic system, we calculate its dynamical tidal radius ($r_t$) due to the LMC following equation (7) of \citet{1983_equation}, as applied in \citet{SMASH1}:
\begin{equation*}
r_{\rm t} \simeq 0.5 \left(\frac{ M_{\rm DELVE 2}}{ M_{\rm LMC} (D_{\rm LMC})} \right)^{1/3}D_{\rm LMC},
\label{eqn:tidal}
\end{equation*}

\noindent where $D_{\text{LMC}}$ refers to the 3D separation between \stamp and the LMC calculated above, and $M_{\rm LMC} (D_{\rm LMC})$ is the enclosed mass of the LMC within a radius of $D_{\rm LMC}$. 

We find that whether or not \stamp is undergoing tidal disruption from the LMC depends sensitively on its dynamical mass --- a key determinant for the system's classification as an ultra-faint galaxy or star cluster.
In order for the dynamical tidal radius of \stamp to be smaller than $\roughly 4.5r_{\rm h} (\roughly 90 \pc)$, corresponding to the maximum radius at which we find member stars identified with \ugali membership probability greater than $5 \%$ ($p_{\rm \ugali} > 0.05$), we calculate that the upper bound on the system's mass-to-light ratio must be $M/L \lesssim 36$, assuming a value of $1.06 \times 10^{11} \Msun$ as the LMC enclosed mass within 30 kpc \citep{Skymapper:2020}, where this radius corresponds to the approximate 3D separation between \stamp and the LMC. 
\par Therefore, if \stamp is found to be most consistent with the population of known ultra-faint dwarf galaxies, which are known to exhibit a wide range of (large) mass-to-light ratios ($30 \lesssim M/L  \lesssim 1000$; \citealt{McConnachie:2012}), we find that it is somewhat unlikely, but not impossible, that the system is undergoing tidal disruption due to the influence of the LMC. 
While a  mass-to-light ratio of $M/L \sim 36$ is not implausible with respect to the full Milky Way satellite galaxy population, this would be significantly lower than the mass-to-light ratio of the dwarf galaxy satellite Segue~1, which is of comparable absolute visual magnitude to \stamp ($M_{V} = -1.5^{+0.6}_{-0.8}$) but has $M/L  \sim 1320$ \citep{Seg1}.\footnote{While the Willman~1 system has similar $M_V$ to \stamp, the system's dynamical state is unclear \citep{2011AJ....142..128W}. Thus, the spectroscopically derived $M/L$ ratio for that system may be unreliable, and therefore is not ideal for comparison to \stamp.}
\par In contrast, if \stamp is found to have a $M/L$ ratio consistent with the population of known star clusters, which typically exhibit mass-to-light ratios of $M/L$ $\roughly 1-2$ \citep{GCML},  we calculate a tidal radius for the system of $r_{\rm t} \lesssim 36 \pc$. 
This upper bound lies at $ \roughly 1.5r_{\rm h}$ for \stamp and thus is well within the maximum system radius of $\roughly 4.5r_{\rm h}$. 
In fact, approximately half of the stars with $p_{\rm \ugali} > 0.05$ lie outside of this purported dynamical tidal radius.
In this case, it would be almost certain that \stamp is undergoing tidal disruption due to the LMC provided its dynamical properties are found to be consistent with other recently discovered star clusters. Lastly, because these calculations only account for the gravitational influence of the LMC at the position of \stamp, we would expect an even lower estimate of the dynamical tidal radius for the system (in both the cluster and dwarf galaxy cases) when including the influence of the SMC.
\par Evidence of tidal disruption (or lack thereof) could in principle be used to help discern whether or not \stamp is a dark-matter dominated system. 
For example, \citet{Simon_2017} utilized a similar argument and the observed tidal tails of the ultra-faint system Tucana~III to constrain its mass-to-light ratio to be $20 < M/L < 240$, providing tentative evidence for a dark-matter dominated nature of the system, consistent with known dwarf galaxies. However, the lack of any clear morphological signatures of tidal disruption for \stamp makes such a determination difficult based on the photometric and astrometric data alone, and the incomplete coverage near the system makes it impossible to conclusively state the absence of tidal disruption, and thus deeper and more complete imaging would be required to make a conclusion based on photometric properties alone. Additionally, spectroscopic followup may reveal tidal signatures beyond what photometric data alone can reveal, such as a velocity gradient, and thus conclusions about whether this system is undergoing tidal disruption/stripping or mass loss are best left until those data are available.
However, we note that such an approach to the classification of this system could prove especially useful if \stamp is found to have a mass low enough such that the velocity dispersion of the system cannot be resolved due to the systematics floor limiting current spectroscopic instruments, as was the case with Segue~2 \citep{SegII}.
We further discuss methods for the classification of \stamp in the following subsection. 

\subsection{Classification of \stamp}

As is evident in \figref{satellites}, \stamp occupies a regime in the size-luminosity plane  that makes it difficult to definitively classify the system as either an ultra-faint halo star cluster or a dwarf galaxy. The continuum between halo star clusters and ultra-faint dwarfs, roughly corresponding to $M_V \gtrsim -2 \magn$ and $10 \pc \lesssim r_{1/2} \lesssim 40\pc$ has been described in the literature as the ``valley of ambiguity'' or similarly the ``trough of uncertainty" \citep[e.g.,][]{Koposov:2015b,Conn:2018a,Conn:2018b}. While the system appears generally more consistent with the ultra-faint galaxy population, as viewed in \figref{satellites}, few conclusions about the true nature of the system can be made in the absence of velocity and metallicity information. 
\par In absence of this information, comparing the morphological and photometric similarities of \stamp to multiple other known systems offers some basic insight into its classification. In particular, \stamp has a similar absolute magnitude to the ultra-faint dwarf galaxies Segue~1, Segue~2, Willman~1, Bootes~II,  Carina~III ($M_V = -1.3, -2.08, -2.53, -2.94, -2.4$, respectively; \citealt{2018ApJ...860...66M, Torrealba:2018a}). While these satellites are all slightly larger than \stamp in terms of major-axis length ($a_{1/2} = 24.2, 38.3, 27.7, 37.3, 30 \pc$, respectively) and more elliptical ($\ellip = 0.33, 0.22,0.47, 0.25 ,0.55$), the azimuthally averaged radii of these systems are close to that of \stamp ($r_{1/2} = 19.8, 33.8, 20.2, 32.3, 20.1 \pc$). 
\par The Milky Way halo star cluster with properties most similar to \stamp (seen as an open red circle directly below the gold star in \figref{satellites}) is DES J0225+0304.
This cluster system was likely tidally stripped from the Sagittarius dwarf galaxy \citep{DES2}, and thus the similarity between \stamp and this system in terms of absolute magnitude and size may reflect analogous origins as stripped satellites of a larger galaxy accreting onto the Milky Way.
% \FIXME{ The large ellipticity and extension of this cluster system ($\epsilon = 0.61$, $a_{1/2} = 18.55 \pc$) are believed to be the result of the fact that the system was likely tidally stripped from the Sagittarius dwarf galaxy \citep{DES2}, and thus the similarity between \stamp and this system in terms of absolute magnitude and size may reflect their analogous origins as stripped satellites of a larger galaxy accreting onto the Milky Way.}
Compared to the remaining population of Milky Way halo star clusters, \stamp lies at a slightly fainter surface brightness, $\mu = 28.2$ mag~arcsec$^{-2}$, than most known systems. We compare \stamp to the LMC/SMC cluster population in the following subsection.
\par Lastly, although not sufficient to make a conclusive judgement about the nature of stellar systems in isolation, the ellipticity of ultra-faint systems including \stamp may also offer insight into their morphologies and dynamical states, and thus can contribute to a broader case about their classifications.
Compared to the known population of Milky Way ultra-faint satellite galaxies, \stamp appears to stand out due to its unusually low ellipticity, even if the upper bound ellipticity of $\epsilon < 0.18$ is considered. In general, globular clusters and their ultra-faint cluster analogs display lower ellipticity compared to their dwarf galaxy counterparts \citep[e.g.,][]{Harris:1996,McConnachie:2012}.
Even though our modelling and parameter fit account for the incomplete coverage near \stamp, judgements based on ellipticity are best reserved for  deeper, more complete imaging. 
Deep imaging of this system could provide concrete assessment of photometric completeness and better disentangle member stars from foreground and background objects \citep[e.g.,][]{Mutlu-Pakdil:2018,Mutlu-Pakdil:2019}. We also note that this system's major axis could be aligned with the line--of--sight, and thus it is possible that the system's ellipticity is not well represented by the projected (2D) distribution of stars displayed in \figref{membership}.
% \citep[e.g., Leo~V;][]{Mutlu-Pakdil:2019}. 
\par Ultimately, the physical classification of \stamp can best be made by measuring its stellar velocity dispersion and estimating its dynamical mass. 
A large velocity dispersion and derived mass-to-light ratio would suggest a dark-matter dominated system, resulting in classification as a (probable) dwarf galaxy, while a smaller measured dispersion might suggest a lower dynamical mass and the absence of dark matter characteristic of the population of outer Milky Way star clusters \citep{Simon:2019}.
Alternately, spectroscopic measurement of a large metallicity dispersion for \stamp could imply the existence of multiple generations of star formation, resulting in a more probable classification of the system as an ultra-faint dwarf galaxy \citep{Willman:2012}.
Such an approach has been applied to a wide variety of ultra-faint systems, and has proved to be a useful discriminant between dwarf galaxies and clusters where it is difficult to resolve a velocity dispersion. This method was notably used by \citet{SegII} for the extremely low-mass galaxy Segue~2, which  appears to be similar to \stamp (with the exception of ellipticity). For Segue~2, only an upper limit could be placed on the velocity dispersion of the system, but the wide spread of metallicities present allowed for a reasonably confident determination that the system is most likely a dwarf galaxy, rather than a star cluster. Such an approach could be applied to reach a conclusive classification of \stamp. 
\par We note that the four bright blue horizontal branch stars and one bright red giant branch star with proper motions consistent with the systemic motion of \stamp are accessible targets for future spectroscopic followup.
In spite of these clear targets, the limited number of identifiable brighter red giant branch member stars available, the compactness of the central core of the system, and the possibility that the velocity dispersion for this (possibly low-mass) system may be at or below the systematic floor of current instruments, may make the classification of this system challenging.

\par If \stamp is confidently identified in follow-up studies as an ultra-faint dwarf galaxy, it should be referred to as Hydrus~II, following the convention that ultra-faint dwarf galaxies are named after the constellation they reside in.
Alternately, if \stamp's properties prove to be most consistent with the population of recently discovered star clusters, the system should continue to be referred to as DELVE 2, following the convention that faint star clusters are named after the survey they are discovered in. Follow-up studies are needed to make a conclusion about the classification of \stamp, and its true nature remains unclear at this time. 
\subsection{Comparison to Known LMC/SMC Cluster Systems}
Given the likelihood that \stamp is associated with the Magellanic Clouds, it is worth probing the origins of this candidate system and its relationship to the known population of LMC/SMC star clusters.
\par The LMC is known to have brought a large population of star clusters as it has been accreted onto the Milky Way \citep{Bica:2008}, and thus it is possible that \stamp may share a similar history to the thousands of known star clusters in the Magellanic system. 
The LMC star cluster formation history is believed to be three-staged, including a period of rapid cluster star formation in the early Universe ($\tau \gtrsim 10$ Gyr), followed by a long quiescent period between $\roughly 10$ Gyr and $\roughly 2-4$ Gyr ago, and then by a period of rapid star cluster formation extending to the present day, potentially due to the interaction between the LMC and SMC \citep{2009AJ....138.1243H, LMCSFH_1, LMCSFH_2,LMCSFH_3}.
One consequence of this period of quiescence in the LMC cluster formation history is the so-called ``age gap" in the age distribution of LMC clusters, with a small ($N \lesssim 20$) population of globular clusters with ages comparable to most known Milky Way globular clusters, separated by the gap from a much larger population of less massive young clusters \citep[e.g.,][]{1992ApJ...388..400B, 1995A&A...298...87G, 1996ARA&A..34..511O}.
These two populations of clusters obey an overarching age-metallicity relation, within which the older clusters ($ \tau > 12 \Gyr$) are significantly more metal poor ($ -2.2 \lesssim [\rm Fe/H] \lesssim −1.2$) compared to the younger population of clusters ($[\rm Fe/H] \gtrsim  −0.7$) \citep{2014MNRAS.438.1067M}.\footnote{We note that \citet{2020arXiv200700341G} recently discovered 16 cluster candidates believed to be within the LMC cluster age gap ($ 4 \Gyr \lesssim \tau \lesssim 10 \Gyr$).}
Therefore, although the photometrically derived metallicity and age for \stamp is limited in accuracy by the small number of red-giant-branch stars available to precisely constrain these properties through synthetic isochrone fitting, it is clear that \stamp is more consistent with an old, metal-poor stellar population and thus the former class of LMC clusters (provided the system is not a dwarf galaxy, as discussed in the previous subsection).
\par While the age and metallicity of \stamp appear to be consistent with the older cluster population of the LMC described above, the position, absolute magnitude, and surface brightness differ significantly from the bright, massive ``classical" globular clusters associated with the early epoch of LMC cluster formation (similar to those of the Milky Way; \figref{satellites}). 
One potential explanation for the significant divergence between these old LMC star clusters and \stamp could be that \stamp originated as a more massive globular cluster associated with the early epoch of LMC star formation, but has been tidally stripped over time due to tidal forces associated with the LMC's interactions with the SMC in the last 4 Gyr.
\par Under this scenario, the observed system at present may be the remaining compact core of an older system, which would explain the apparent incompatability of the \stamp's morphological properties with those of Galactic/LMC globulars with similar age/metallicity. However, given its relatively large current distance from the LMC/SMC system (compared to known cluster systems), such extreme mass loss may only have been possible if \stamp's orbit is highly elliptical, with a pericenter extremely close to the LMC or SMC. In such an eccentric orbit, several mechanisms have been proposed to contribute to high degrees of cluster mass loss, including enhanced evaporation due to increased tidal heating in the varying tidal field \citep{10.1093/mnras/staa2425}, the stripping of energetic stars at the cluster outskirts due to the shrinking of the cluster Jacobi radius at its orbital pericenter \citep{10.1093/mnras/stu961}, and tidal shocks induced either when a cluster's stars have orbital periods longer than the pericenter interaction time between the cluster and the host galaxy \citep{10.1093/mnras/staa2425} or due to the cluster's interaction with other types of astrophysical systems \citep[e.g., molecular clouds;][]{10.1093/mnrasl/slw163}.
However, without radial velocity information or any visible signs of tidal disruption/mass loss in the current DELVE data, such a conclusion about \stamp's orbital history is difficult to test.

\par In contrast to the conclusions reached above for the LMC, the old age of \stamp appears inconsistent with the known population of SMC star clusters. Unlike the LMC, the SMC is known to host a population of globular clusters with an almost continuous distribution in age up to $\tau \sim 8$ Gyr \citep[e.g.,][]{HZ04,Dias2010}. Only one SMC cluster, NGC121, serves as an exception to this distribution, with an age of $\tau \sim 11$ Gyr, making the system the oldest known SMC cluster \citep{121age}. This system has a metallicity of $[\rm Fe/H] \sim −1.28$ \citep{121ref2}.
While our metallicity and age estimates for \stamp have fairly large uncertainties, the values we derived appear to be  older and more metal poor than NGC121 --- currently accepted to be the only old, Milky-Way-like globular cluster in the SMC \citep{121age} --- and thus, it is unlikely to have an SMC origin, unless this system is unique in its age/metallicity or represents a new class of previously undiscovered objects. 
With 6D motion information for this system, along with information about its chemical abundances, it may be possible in the future to probe these distinct theories for \stamp, and in doing so reach a comprehensive conclusion about the system's origin, evolution, and classification.

\section{Summary}
\label{sec:conclusion}
We have presented the discovery of an ultra-faint resolved stellar system, \Doublestamp, in a search of $\roughly 2,200 \deg^{2}$ of early data from the DELVE survey in the Magellanic periphery, representing the third ultra-faint system discovered by the DELVE project to date. 
This new ultra-faint stellar system was detected at high confidence by a search for spatial overdensities of stars consistent with an old, metal-poor stellar population.
%In particular, after searching $\roughly 2,200 \deg^{2}$ for spatial overdensities of stars consistent with a template isochrone corresponding to an old, metal-poor stellar population, we identified this new ultra-faint stellar system at high confidence.  
Based on maximum likelihood fits to the system's morphological and isochrone properties alone, we found that the system is consistent with an old, metal-poor stellar population, and by utilizing proper motions from \Gaia DR2, we tentatively confirmed that \stamp appears to be a gravitationally bound association of stars with coherent motion on the sky, and found that the spatial position, distance, and proper motion strongly suggest an association between \stamp and the LMC/SMC. However, we were unable to draw a robust conclusion about whether the system is more consistent with a dark-matter-dominated dwarf galaxy or a faint star cluster. With three new satellites now identified by DELVE, and with numerical simulations predicting that $\roughly 100$ Milky Way satellites with $M_V < 0\magn$ and $r_{1/2} > 10\pc$ remaining to be discovered \citep{Nadler2020}, we anticipate DELVE will continue to play an important role in advancing our understanding of the Milky Way satellite system as the survey continues its comprehensive census of the southern sky.

%-------------------------------------------------------------------------------

\section{Acknowledgments}
WC acknowledges support from the University of Chicago Liew Family Research Fellowship. This work was supported in part by the U.S. Department of Energy, Office of Science, Office of Workforce Development for Teachers and Scientists (WDTS) under the Science Undergraduate Laboratory Internships (SULI) program. 
The DELVE project is partially supported by Fermilab LDRD project L2019-011 and the NASA Fermi Guest Investigator Program Cycle 9 No. 91201.
ABP acknowledges support from NSF grant AST-1813881.
This research received support from the National Science Foundation (NSF) under grant no. NSF DGE-1656518 through the NSF Graduate Research Fellowship received by S.M.
JLC acknowledges support from NSF grant AST-1816196.
JDS acknowledges support from NSF grant AST-1714873.
SRM acknowledges support from NSF grant AST-1909497.
DMD acknowledges financial support from the State Agency for Research
of the Spanish MCIU through the ``Centre of Excellence Severo Ochoa''
award for the Instituto de Astrofísica de
Andalucía (SEV-2017-0709).

This project used data obtained with the Dark Energy Camera (DECam), which was constructed by the Dark Energy Survey (DES) collaboration.
Funding for the DES Projects has been provided by 
the DOE and NSF (USA),   
MISE (Spain),   
STFC (UK), 
HEFCE (UK), 
NCSA (UIUC), 
KICP (U. Chicago), 
CCAPP (Ohio State), 
MIFPA (Texas A\&M University),  
CNPQ, 
FAPERJ, 
FINEP (Brazil), 
MINECO (Spain), 
DFG (Germany), 
and the collaborating institutions in the Dark Energy Survey, which are
Argonne Lab, 
UC Santa Cruz, 
University of Cambridge, 
CIEMAT-Madrid, 
University of Chicago, 
University College London, 
DES-Brazil Consortium, 
University of Edinburgh, 
ETH Z{\"u}rich, 
Fermilab, 
University of Illinois, 
ICE (IEEC-CSIC), 
IFAE Barcelona, 
Lawrence Berkeley Lab, 
LMU M{\"u}nchen, and the associated Excellence Cluster Universe, 
University of Michigan, 
NSF's National Optical-Infrared Astronomy Research Laboratory, 
University of Nottingham, 
Ohio State University, 
OzDES Membership Consortium
University of Pennsylvania, 
University of Portsmouth, 
SLAC National Lab, 
Stanford University, 
University of Sussex, 
and Texas A\&M University.

This work has made use of data from the European Space Agency (ESA) mission {\it Gaia} (\url{https://www.cosmos.esa.int/gaia}), processed by the {\it Gaia} Data Processing and Analysis Consortium (DPAC, \url{https://www.cosmos.esa.int/web/gaia/dpac/consortium}).
Funding for the DPAC has been provided by national institutions, in particular the institutions participating in the {\it Gaia} Multilateral Agreement.

Based on observations at Cerro Tololo Inter-American Observatory, NSF's National Optical-Infrared Astronomy Research Laboratory (2019A-0305; PI: Drlica-Wagner), which is operated by the Association of Universities for Research in Astronomy (AURA) under a cooperative agreement with the National Science Foundation.

This manuscript has been authored by Fermi Research Alliance, LLC, under contract No.\ DE-AC02-07CH11359 with the US Department of Energy, Office of Science, Office of High Energy Physics. The United States Government retains and the publisher, by accepting the article for publication, acknowledges that the United States Government retains a non-exclusive, paid-up, irrevocable, worldwide license to publish or reproduce the published form of this manuscript, or allow others to do so, for United States Government purposes.

\facility{Blanco, \Gaia.}
\software{\code{astropy} \citep{astropy:2013,astropy:2018}, \emcee \citep{Foreman-Mackey:2013}, \code{fitsio},\footnote{\url{https://github.com/esheldon/fitsio}} \healpix \citep{Gorski:2005},\footnote{\url{http://healpix.sourceforge.net}} \code{healpy},\footnote{\url{https://github.com/healpy/healpy}} \code{Matplotlib} \citep{Hunter:2007}, \code{numpy} \citep{numpy:2011}, \code{scipy} \citep{scipy:2001}, \ugali \citep{Bechtol:2015}.\footnote{\url{https://github.com/DarkEnergySurvey/ugali}}}

\bibliography{main}

\end{document}

%% file: commands.tex
\usepackage{amsmath}
\usepackage{amssymb}
\usepackage{xspace}
\usepackage{xifthen}
\usepackage{eso-pic}

\newcommand{\Gaia}{{\it Gaia}\xspace}

\definecolor{forestgreen}{HTML}{228B22}
\definecolor{urlblue}{HTML}{000000}

\mathchardef\mhyphen="2D

\newcommand{\roughly}{\ensuremath{ {\sim}\,} }

\newlength{\dhatheight}

\newcommand{\code}[1]{\texttt{#1}\xspace}

\newcommand{\unit}[1]{\ensuremath{\mathrm{\,#1}}\xspace}
\newcommand{\yr}{\unit{yr}}
\newcommand{\Gyr}{\unit{Gyr}}

\newcommand{\degree}{\ensuremath{{}^{\circ}}\xspace}

\newcommand{\mas}{\unit{mas}}

\newcommand{\pc}{\unit{pc}}
\newcommand{\kpc}{\unit{kpc}}

\newcommand{\Msun}{\unit{M_\odot}}

\newcommand{\magn}{\unit{mag}}

\newcommand{\secref}[1]{Section~\ref{sec:#1}}

\newcommand{\tabref}[1]{Table~\ref{tab:#1}}

\newcommand{\figref}[1]{Figure~\ref{fig:#1}}

\newcommand{\bandvar}[2][]{%
  \ifthenelse{\isempty{#1}}{\var{#2}}{\var{#2\_#1}}%
}

\newcommand{\modulus}{\ensuremath{m - M}\xspace}

\newcommand{\ra}{{\ensuremath{\alpha_{2000}}}\xspace}
\newcommand{\dec}{{\ensuremath{\delta_{2000}}}\xspace}
\newcommand{\age}{{\ensuremath{\tau}}\xspace}
\newcommand{\metal}{{\ensuremath{Z}}\xspace}

\newcommand{\feh}{{\ensuremath{\rm [Fe/H]}}\xspace}
\newcommand{\ellip}{\ensuremath{\epsilon}\xspace}
\newcommand{\PA}{\ensuremath{\mathrm{P.A.}}\xspace}
\newcommand{\TS}{\ensuremath{\mathrm{TS}}\xspace}

\newcommand{\HEALPix}{\code{HEALPix}}
\newcommand{\healpix}{\HEALPix}

\newcommand{\emcee}{\code{emcee}}
\newcommand{\ugali}{\code{ugali}}
\newcommand{\simple}{\code{simple}}
\newcommand{\var}[1]{\ensuremath{\texttt{\MakeUppercase{#1}}}\xspace}
\newcommand{\nside}{\code{nside}}

\providecommand\physrep{\ref@jnl{Phys.~Rep.}}%
\providecommand\apjs{\ref@jnl{ApJS}}%
\providecommand{\jcap}{\ref@jnl{JCAP}}%

%% file: authors.tex
% Author list file generated with: mkauthlist 1.2.3 
% mkauthlist authors.csv 

\author[0000-0003-1697-7062]{W.~Cerny}

\affiliation{Kavli Institute for Cosmological Physics, University of Chicago, Chicago, IL 60637, USA}
\affiliation{Department of Astronomy and Astrophysics, University of Chicago, Chicago IL 60637, USA}
\affiliation{Fermi National Accelerator Laboratory, P.O.\ Box 500, Batavia, IL 60510, USA}
\author[0000-0002-6021-8760]{A.~B.~Pace}
\affiliation{McWilliams Center for Cosmology, Carnegie Mellon University, 5000 Forbes Avenue, Pittsburgh, PA 15213, USA}

\author[0000-0001-8251-933X]{A.~Drlica-Wagner}
\affiliation{Fermi National Accelerator Laboratory, P.O.\ Box 500, Batavia, IL 60510, USA}
\affiliation{Kavli Institute for Cosmological Physics, University of Chicago, Chicago, IL 60637, USA}
\affiliation{Department of Astronomy and Astrophysics, University of Chicago, Chicago IL 60637, USA}

\author[0000-0001-6957-1627]{P.~S.~Ferguson}
\affiliation{George P. and Cynthia Woods Mitchell Institute for Fundamental Physics and Astronomy, Texas A\&M University, College Station, TX 77843, USA}
\affiliation{Department of Physics and Astronomy, Texas A\&M University, College Station, TX 77843, USA}

\author[0000-0003-3519-4004]{S.~Mau}
% \affiliation{Kavli Institute for Cosmological Physics, University of Chicago, Chicago, IL 60637, USA}
\affiliation{Department of Physics, Stanford University, 382 Via Pueblo Mall, Stanford, CA 94305, USA}
\affiliation{Kavli Institute for Particle Astrophysics \& Cosmology, P.O.\ Box 2450, Stanford University, Stanford, CA 94305, USA}

% Old Author List: 
\author[0000-0002-6904-359X]{M.~Adam\'ow}
\affiliation{National Center for Supercomputing Applications, University of Illinois, 1205 West Clark Street, Urbana, IL 61801, USA}
\author[0000-0002-3936-9628]{J.~L.~Carlin}
\affiliation{Rubin Observatory/AURA, 950 North Cherry Avenue, Tucson, AZ, 85719, USA}

\author[0000-0003-1680-1884]{Y.~Choi}
\affiliation{Space Telescope Science Institute, 3700 San Martin Drive, Baltimore, MD 21218, USA}

\author[0000-0002-8448-5505]{D.~Erkal}
\affiliation{Department of Physics, University of Surrey, Guildford GU2 7XH, UK}

%%%%%%%%%%%%%%%%%%%%%%%%%%%%%%%%%%%%%%%%%%%%%%%55

% \author[0000-0002-4588-6517]{R.~A.~Gruendl}
% \affiliation{Department of Astronomy, University of Illinois, 1002 W. Green Street, Urbana, IL 61801, USA}
% \affiliation{National Center for Supercomputing Applications, University of Illinois, 1205 West Clark Street, Urbana, IL 61801, USA}
% \author{D.~Hernandez-Lang}
% \affiliation{University of La Serena, La Serena, Chile}
% \affiliation{Cerro Tololo Inter-American Observatory, NSF's National Optical-Infrared Astronomy Research Laboratory, Casilla 603, La Serena, Chile}
% \affiliation{Gemini Observatory, La Serena, Chile}

\author[0000-0001-6421-0953]{L.~C.~Johnson}
\affiliation{Center for Interdisciplinary Exploration and Research in Astrophysics (CIERA) and Department of Physics and Astronomy, Northwestern University, 1800 Sherman Ave, Evanston, IL 60201 USA}

\author[0000-0002-9110-6163]{T.~S.~Li}
\altaffiliation{NHFP Einstein Fellow}
\affiliation{Observatories of the Carnegie Institution for Science, 813 Santa Barbara Street, Pasadena, CA 91101, USA}
\affiliation{Department of Astrophysical Sciences, Princeton University, Princeton, NJ 08544, USA}
\author[0000-0002-9144-7726]{C.~E.~Mart\'inez-V\'azquez}
\affiliation{Cerro Tololo Inter-American Observatory, NSF's National Optical-Infrared Astronomy Research Laboratory, Casilla 603, La Serena, Chile}
% \author{E.~Morganson}
% \affiliation{National Center for Supercomputing Applications, University of Illinois, 1205 West Clark Street, Urbana, IL 61801, USA}
\author[0000-0001-9649-4815]{B.~Mutlu-Pakdil}
\affiliation{Kavli Institute for Cosmological Physics, University of Chicago, Chicago, IL 60637, USA}
\affiliation{Department of Astronomy and Astrophysics, University of Chicago, Chicago IL 60637, USA}
% \affiliation{Department of Astronomy/Steward Observatory, 933 North Cherry Avenue, Room N204, Tucson, AZ 85721-0065, USA}

\author[0000-0002-1793-3689]{D.~L.~Nidever}
\affiliation{Department of Physics, Montana State University, P.O. Box 173840, Bozeman, MT 59717-3840, USA}
\affiliation{NSF's National Optical-Infrared Astronomy Research Laboratory, 950 N. Cherry Ave., Tucson, AZ 85719, USA}
\author[0000-0002-7134-8296]{K.~A.~G.~Olsen}
\affiliation{NSF's National Optical-Infrared Astronomy Research Laboratory, 950 N. Cherry Ave., Tucson, AZ 85719, USA}

\author[0000-0001-9186-6042]{A.~Pieres}
\affiliation{Laborat\'orio Interinstitucional de e-Astronomia - LIneA, Rua Gal. Jos\'e Cristino 77, Rio de Janeiro, RJ - 20921-400, Brazil }
\affiliation{Observat\'orio Nacional, Rua Gal. Jos\'e Cristino 77, Rio de Janeiro, RJ - 20921-400, Brazil}

\author[0000-0002-9599-310X]{E.~J.~Tollerud}
\affiliation{Space Telescope Science Institute, 3700 San Martin Drive, Baltimore, MD 21218, USA}

\author{J.~D.~Simon}
\affiliation{Observatories of the Carnegie Institution for Science, 813 Santa Barbara Street, Pasadena, CA 91101, USA}

\author[0000-0003-4341-6172]{A.~K.~Vivas}
\affiliation{Cerro Tololo Inter-American Observatory, NSF's National Optical-Infrared Astronomy Research Laboratory, Casilla 603, La Serena, Chile}

%%%%%%%%%%%%%%%%%%%%%%%%%%%%%%%%%%%%%%%%%%%%
\author[0000-0001-5160-4486]{D.~J.~James}
% \affiliation{Center for Astrophysics, Harvard \& Smithsonian, 60 Garden Street, Cambridge, MA 02138, USA}
\affiliation{ASTRAVEO, LLC, PO Box 1668, Gloucester, MA 01931}

\author[0000-0003-2511-0946]{N.~Kuropatkin}
\affiliation{Fermi National Accelerator Laboratory, P.O.\ Box 500, Batavia, IL 60510, USA}

\author[0000-0003-2025-3147]{S.~Majewski}
\affiliation{Department of Astronomy, University of Virginia, Charlottesville, VA, 22904, USA}

\author{D.~Mart\'{i}nez-Delgado}
\affiliation{Instituto de Astrof\'{i}sica de Andaluc\'{i}a, CSIC, E-18080 Granada, Spain}

\author[0000-0002-8093-7471]{P.~Massana}
\affiliation{Department of Physics, University of Surrey, Guildford GU2 7XH, UK}

\author{A.~Miller}
\affiliation{Leibniz-Institut für Astrophysik Potsdam (AIP), An der Sternwarte 16, D-14482 Potsdam, Germany}

\author[0000-0002-7357-0317]{E.~H.~Neilsen}
\affiliation{Fermi National Accelerator Laboratory, P.O.\ Box 500, Batavia, IL 60510, USA}

\author{N.~E.~D.~No\"el}
\affiliation{Department of Physics, University of Surrey, Guildford GU2 7XH, UK}

\author[0000-0001-5805-5766]{A.~H.~Riley}
\affiliation{George P. and Cynthia Woods Mitchell Institute for Fundamental Physics and Astronomy, Texas A\&M University, College Station, TX 77843, USA}
\affiliation{Department of Physics and Astronomy, Texas A\&M University, College Station, TX 77843, USA}

\author[0000-0003-4102-380X]{D.~J.~Sand}
\affiliation{Department of Astronomy/Steward Observatory, 933 North Cherry Avenue, Room N204, Tucson, AZ 85721-0065, USA}

\author[0000-0003-3402-6164]{L.~Santana-Silva}
\affiliation{NAT-Universidade Cruzeiro do Sul / Universidade Cidade de S{\~a}o Paulo, Rua Galv{\~a}o Bueno, 868, 01506-000, S{\~a}o Paulo, SP, Brazil}

\author[0000-0003-1479-3059]{G.~S.~Stringfellow}
\affiliation{Center for Astrophysics and Space Astronomy, University of Colorado, 389 UCB, Boulder, CO 80309-0389, USA}

\author{D.~L.~Tucker}
\affiliation{Fermi National Accelerator Laboratory, P.O.\ Box 500, Batavia, IL 60510, USA}

\correspondingauthor{William Cerny, Andrew B. Pace}
\email{williamcerny@uchicago.edu, apace@andrew.cmu.edu}

% \color{red}
\collaboration{(DELVE Collaboration)}
% \color{black}

%% file: main.bbl
\begin{thebibliography}{}
\expandafter\ifx\csname natexlab\endcsname\relax\def\natexlab#1{#1}\fi
\providecommand{\url}[1]{\href{#1}{#1}}

\bibitem[{{Astropy Collaboration} {et~al.}(2013){Astropy Collaboration},
  {Robitaille}, {Tollerud}, {Greenfield}, {Droettboom}, {Bray}, {Aldcroft},
  {Davis}, {Ginsburg}, {Price-Whelan}, {Kerzendorf}, {Conley}, {Crighton},
  {Barbary}, {Muna}, {Ferguson}, {Grollier}, {Parikh}, {Nair}, {Unther},
  {Deil}, {Woillez}, {Conseil}, {Kramer}, {Turner}, {Singer}, {Fox}, {Weaver},
  {Zabalza}, {Edwards}, {Azalee Bostroem}, {Burke}, {Casey}, {Crawford},
  {Dencheva}, {Ely}, {Jenness}, {Labrie}, {Lim}, {Pierfederici}, {Pontzen},
  {Ptak}, {Refsdal}, {Servillat}, \& {Streicher}}]{astropy:2013}
{Astropy Collaboration}, {Robitaille}, T.~P., {Tollerud}, E.~J., {et~al.} 2013,
  \aap, 558, A33

\bibitem[{Balbinot {et~al.}(2013)Balbinot, Santiago, da~Costa, Maia, Majewski,
  Nidever, Rocha-Pinto, Thomas, Wechsler, \& Yanny}]{Balbinot:2013}
Balbinot, E., Santiago, B.~X., da~Costa, L., {et~al.} 2013, \apj, 767, 101.
\newblock \url{http://stacks.iop.org/0004-637X/767/i=2/a=101}

\bibitem[{Bechtol {et~al.}(2015)Bechtol, Drlica-Wagner, Balbinot,
  {et~al.}}]{Bechtol:2015}
Bechtol, K., Drlica-Wagner, A., Balbinot, E., {et~al.} 2015, \apj, 807, 50

\bibitem[{{Belokurov} {et~al.}(2014){Belokurov}, {Irwin}, {Koposov}, {Evans},
  {Gonzalez-Solares}, {Metcalfe}, \& {Shanks}}]{Belokurov:2014}
{Belokurov}, V., {Irwin}, M.~J., {Koposov}, S.~E., {et~al.} 2014, \mnras, 441,
  2124

\bibitem[{{Belokurov} {et~al.}(2006){Belokurov}, {Zucker}, {Evans},
  {Wilkinson}, {Irwin}, {Hodgkin}, {Bramich}, {Irwin}, {Gilmore}, {Willman},
  {Vidrih}, {Newberg}, {Wyse}, {Fellhauer}, {Hewett}, {Cole}, {Bell}, {Beers},
  {Rockosi}, {Yanny}, {Grebel}, {Schneider}, {Lupton}, {Barentine},
  {Brewington}, {Brinkmann}, {Harvanek}, {Kleinman}, {Krzesinski}, {Long},
  {Nitta}, {Smith}, \& {Snedden}}]{Belokurov2006ApJ...647L.111B}
{Belokurov}, V., {Zucker}, D.~B., {Evans}, N.~W., {et~al.} 2006, \apjl, 647,
  L111

\bibitem[{{Belokurov} {et~al.}(2007){Belokurov}, {Zucker}, {Evans}, {Kleyna},
  {Koposov}, {Hodgkin}, {Irwin}, {Gilmore}, {Wilkinson}, {Fellhauer},
  {Bramich}, {Hewett}, {Vidrih}, {De Jong}, {Smith}, {Rix}, {Bell}, {Wyse},
  {Newberg}, {Mayeur}, {Yanny}, {Rockosi}, {Gnedin}, {Schneider}, {Beers},
  {Barentine}, {Brewington}, {Brinkmann}, {Harvanek}, {Kleinman}, {Krzesinski},
  {Long}, {Nitta}, \& {Snedden}}]{Belokurov2007ApJ...654..897B}
---. 2007, \apj, 654, 897

\bibitem[{{Belokurov} {et~al.}(2009){Belokurov}, {Walker}, {Evans}, {Gilmore},
  {Irwin}, {Mateo}, {Mayer}, {Olszewski}, {Bechtold}, \&
  {Pickering}}]{Belokurov2009MNRAS.397.1748B}
{Belokurov}, V., {Walker}, M.~G., {Evans}, N.~W., {et~al.} 2009, \mnras, 397,
  1748

\bibitem[{{Belokurov} {et~al.}(2010){Belokurov}, {Walker}, {Evans}, {Gilmore},
  {Irwin}, {Just}, {Koposov}, {Mateo}, {Olszewski}, {Watkins}, \&
  {Wyrzykowski}}]{2010ApJ...712L.103B}
---. 2010, \apjl, 712, L103

\bibitem[{{Bernstein} {et~al.}(2018){Bernstein}, {Abbott}, {Armstrong},
  {Burke}, {Diehl}, {Gruendl}, {Johnson}, {Li}, {Rykoff}, {Walker}, {Wester},
  \& {Yanny}}]{Bernstein:2018}
{Bernstein}, G.~M., {Abbott}, T.~M.~C., {Armstrong}, R., {et~al.} 2018, \pasp,
  130, 054501

\bibitem[{{Bertelli} {et~al.}(1992){Bertelli}, {Mateo}, {Chiosi}, \&
  {Bressan}}]{1992ApJ...388..400B}
{Bertelli}, G., {Mateo}, M., {Chiosi}, C., \& {Bressan}, A. 1992, \apj, 388,
  400

\bibitem[{{Bertin}(2011)}]{Bertin:2011}
{Bertin}, E. 2011, in Astronomical Society of the Pacific Conference Series,
  Vol. 442, Astronomical Data Analysis Software and Systems XX, ed. I.~N.
  {Evans}, A.~{Accomazzi}, D.~J. {Mink}, \& A.~H. {Rots}, 435

\bibitem[{{Bertin} \& {Arnouts}(1996)}]{Bertin:1996}
{Bertin}, E., \& {Arnouts}, S. 1996, \aaps, 117, 393

\bibitem[{{Bica} {et~al.}(2008){Bica}, {Bonatto}, {Dutra}, \&
  {Santos}}]{Bica:2008}
{Bica}, E., {Bonatto}, C., {Dutra}, C.~M., \& {Santos}, J.~F.~C. 2008, \mnras,
  389, 678

\bibitem[{{Bica} {et~al.}(2020){Bica}, {Westera}, {Kerber}, {Dias}, {Maia},
  {Santos}, {Barbuy}, \& {Oliveira}}]{Bica2020}
{Bica}, E., {Westera}, P., {Kerber}, L. d.~O., {et~al.} 2020, \aj, 159, 82

\bibitem[{{Bovy}(2015)}]{galpy}
{Bovy}, J. 2015, \apjs, 216, 29

\bibitem[{{Bressan} {et~al.}(2012){Bressan}, {Marigo}, {Girardi}, {Salasnich},
  {Dal Cero}, {Rubele}, \& {Nanni}}]{Bressan:2012}
{Bressan}, A., {Marigo}, P., {Girardi}, L., {et~al.} 2012, \mnras, 427, 127

\bibitem[{{Burke} {et~al.}(2018){Burke}, {Rykoff}, {Allam}, {Annis}, {Bechtol},
  {Bernstein}, {Drlica-Wagner}, {Finley}, {Gruendl}, {James}, {Kent},
  {Kessler}, {Kuhlmann}, {Lasker}, {Li}, {Scolnic}, {Smith}, {Tucker},
  {Wester}, {Yanny}, {Abbott}, {Abdalla}, {Benoit-L{\'e}vy}, {Bertin}, {Carnero
  Rosell}, {Carrasco Kind}, {Carretero}, {Cunha}, {D'Andrea}, {da Costa},
  {Desai}, {Diehl}, {Doel}, {Estrada}, {Garc{\'\i}a-Bellido}, {Gruen},
  {Gutierrez}, {Honscheid}, {Kuehn}, {Kuropatkin}, {Maia}, {March}, {Marshall},
  {Melchior}, {Menanteau}, {Miquel}, {Plazas}, {Sako}, {Sanchez}, {Scarpine},
  {Schindler}, {Sevilla-Noarbe}, {Smith}, {Smith}, {Soares-Santos}, {Sobreira},
  {Suchyta}, {Tarle}, {Walker}, \& {DES Collaboration}}]{Burke:2018}
{Burke}, D.~L., {Rykoff}, E.~S., {Allam}, S., {et~al.} 2018, \aj, 155, 41

\bibitem[{{Cannon} {et~al.}(1977){Cannon}, {Hawarden}, \&
  {Tritton}}]{Cannon:1977}
{Cannon}, R.~D., {Hawarden}, T.~G., \& {Tritton}, S.~B. 1977, \mnras, 180, 81P

\bibitem[{{Chabrier}(2001)}]{Chabrier:2001}
{Chabrier}, G. 2001, \apj, 554, 1274

\bibitem[{{Chambers} {et~al.}(2016){Chambers}, {Magnier}, {Metcalfe},
  {Flewelling}, {Huber}, {Waters}, {Denneau}, {Draper}, {Farrow}, {Finkbeiner},
  {Holmberg}, {Koppenhoefer}, {Price}, {Saglia}, {Schlafly}, {Smartt},
  {Sweeney}, {Wainscoat}, {Burgett}, {Grav}, {Heasley}, {Hodapp}, {Jedicke},
  {Kaiser}, {Kudritzki}, {Luppino}, {Lupton}, {Monet}, {Morgan}, {Onaka},
  {Stubbs}, {Tonry}, {Banados}, {Bell}, {Bender}, {Bernard}, {Botticella},
  {Casertano}, {Chastel}, {Chen}, {Chen}, {Cole}, {Deacon}, {Frenk},
  {Fitzsimmons}, {Gezari}, {Goessl}, {Goggia}, {Goldman}, {Grebel}, {Hambly},
  {Hasinger}, {Heavens}, {Heckman}, {Henderson}, {Henning}, {Holman}, {Hopp},
  {Ip}, {Isani}, {Keyes}, {Koekemoer}, {Kotak}, {Long}, {Lucey}, {Liu},
  {Martin}, {McLean}, {Morganson}, {Murphy}, {Nieto-Santisteban}, {Norberg},
  {Peacock}, {Pier}, {Postman}, {Primak}, {Rae}, {Rest}, {Riess}, {Riffeser},
  {Rix}, {Roser}, {Schilbach}, {Schultz}, {Scolnic}, {Szalay}, {Seitz},
  {Shiao}, {Small}, {Smith}, {Soderblom}, {Taylor}, {Thakar}, {Thiel},
  {Thilker}, {Urata}, {Valenti}, {Walter}, {Watters}, {Werner}, {White},
  {Wood-Vasey}, \& {Wyse}}]{Chambers:2016}
{Chambers}, K.~C., {Magnier}, E.~A., {Metcalfe}, N., {et~al.} 2016, ArXiv
  e-prints, arXiv:1612.05560

\bibitem[{{Clementini} {et~al.}(2019){Clementini}, {Ripepi}, {Molinaro},
  {Garofalo}, {Muraveva}, {Rimoldini}, {Guy}, {Jevardat de Fombelle},
  {Nienartowicz}, {Marchal}, {Audard}, {Holl}, {Leccia}, {Marconi}, {Musella},
  {Mowlavi}, {Lecoeur-Taibi}, {Eyer}, {De Ridder}, {Regibo}, {Sarro},
  {Szabados}, {Evans}, \& {Riello}}]{Clementini2019A&A...622A..60C}
{Clementini}, G., {Ripepi}, V., {Molinaro}, R., {et~al.} 2019, \aap, 622, A60

\bibitem[{Conn {et~al.}(2018{\natexlab{a}})Conn, Jerjen, Kim, \&
  Schirmer}]{Conn:2018a}
Conn, B.~C., Jerjen, H., Kim, D., \& Schirmer, M. 2018{\natexlab{a}}, The
  Astrophysical Journal, 852, 68.
\newblock \url{http://dx.doi.org/10.3847/1538-4357/aa9eda}

\bibitem[{Conn {et~al.}(2018{\natexlab{b}})Conn, Jerjen, Kim, \&
  Schirmer}]{Conn:2018b}
---. 2018{\natexlab{b}}, The Astrophysical Journal, 857, 70.
\newblock \url{http://dx.doi.org/10.3847/1538-4357/aab61c}

\bibitem[{{Dalessandro} {et~al.}(2016){Dalessandro}, {Lapenna}, {Mucciarelli},
  {Origlia}, {Ferraro}, \& {Lanzoni}}]{121ref2}
{Dalessandro}, E., {Lapenna}, E., {Mucciarelli}, A., {et~al.} 2016, \apj, 829,
  77

\bibitem[{{de Grijs} \& {Bono}(2015)}]{dGB_smc}
{de Grijs}, R., \& {Bono}, G. 2015, \aj, 149, 179

\bibitem[{{de Grijs} {et~al.}(2014){de Grijs}, {Wicker}, \&
  {Bono}}]{Grijs_Distance}
{de Grijs}, R., {Wicker}, J.~E., \& {Bono}, G. 2014, \aj, 147, 122

\bibitem[{{Deason} {et~al.}(2015){Deason}, {Wetzel}, {Garrison-Kimmel}, \&
  {Belokurov}}]{Deason2015MNRAS.453.3568D}
{Deason}, A.~J., {Wetzel}, A.~R., {Garrison-Kimmel}, S., \& {Belokurov}, V.
  2015, \mnras, 453, 3568

\bibitem[{{DES Collaboration} {et~al.}(2005){DES Collaboration}, Abbott,
  Aldering, Annis, {et~al.}}]{Abbott:2005bi}
{DES Collaboration}, Abbott, T., Aldering, G., Annis, J., {et~al.} 2005,
  arXiv:astro-ph/0510346

\bibitem[{{DES Collaboration} {et~al.}(2018){DES Collaboration}, {Abbott},
  {Abdalla}, {Allam}, {et~al.}}]{DR1:2018}
{DES Collaboration}, {Abbott}, T.~M.~C., {Abdalla}, F.~B., {Allam}, S.,
  {et~al.} 2018, \apjs, 239, 18

\bibitem[{{DES Collaboration} {et~al.}(2016){DES Collaboration}, {Abbott},
  {Abdalla}, {Aleksi{\'c}}, {Allam}, {Amara}, {Bacon}, {Balbinot}, {Banerji},
  {Bechtol}, {Benoit-L{\'e}vy}, {Bernstein}, {Bertin}, {Blazek}, {Bonnett},
  {Bridle}, {Brooks}, {Brunner}, {Buckley-Geer}, {Burke}, {Caminha}, {Capozzi},
  {Carlsen}, {Carnero-Rosell}, {Carollo}, {Carrasco-Kind}, {Carretero},
  {Castander}, {Clerkin}, {Collett}, {Conselice}, {Crocce}, {Cunha},
  {D'Andrea}, {da Costa}, {Davis}, {Desai}, {Diehl}, {Dietrich}, {Dodelson},
  {Doel}, {Drlica-Wagner}, {Estrada}, {Etherington}, {Evrard}, {Fabbri},
  {Finley}, {Flaugher}, {Foley}, {Fosalba}, {Frieman}, {Garc{\'\i}a-Bellido},
  {Gaztanaga}, {Gerdes}, {Giannantonio}, {Goldstein}, {Gruen}, {Gruendl},
  {Guarnieri}, {Gutierrez}, {Hartley}, {Honscheid}, {Jain}, {James}, {Jeltema},
  {Jouvel}, {Kessler}, {King}, {Kirk}, {Kron}, {Kuehn}, {Kuropatkin}, {Lahav},
  {Li}, {Lima}, {Lin}, {Maia}, {Makler}, {Manera}, {Maraston}, {Marshall},
  {Martini}, {McMahon}, {Melchior}, {Merson}, {Miller}, {Miquel}, {Mohr},
  {Morice-Atkinson}, {Naidoo}, {Neilsen}, {Nichol}, {Nord}, {Ogando},
  {Ostrovski}, {Palmese}, {Papadopoulos}, {Peiris}, {Peoples}, {Percival},
  {Plazas}, {Reed}, {Refregier}, {Romer}, {Roodman}, {Ross}, {Rozo}, {Rykoff},
  {Sadeh}, {Sako}, {S{\'a}nchez}, {Sanchez}, {Santiago}, {Scarpine},
  {Schubnell}, {Sevilla-Noarbe}, {Sheldon}, {Smith}, {Smith}, {Soares-Santos},
  {Sobreira}, {Soumagnac}, {Suchyta}, {Sullivan}, {Swanson}, {Tarle}, {Thaler},
  {Thomas}, {Thomas}, {Tucker}, {Vieira}, {Vikram}, {Walker}, {Wechsler},
  {Weller}, {Wester}, {Whiteway}, {Wilcox}, {Yanny}, {Zhang}, \&
  {Zuntz}}]{DES:2016}
{DES Collaboration}, {Abbott}, T., {Abdalla}, F.~B., {et~al.} 2016, \mnras,
  460, 1270

\bibitem[{{Desai} {et~al.}(2012){Desai}, {Armstrong}, {Mohr}, {Semler}, {Liu},
  {Bertin}, {Allam}, {Barkhouse}, {Bazin}, {Buckley-Geer}, {Cooper}, {Hansen},
  {High}, {Lin}, {Lin}, {Ngeow}, {Rest}, {Song}, {Tucker}, \&
  {Zenteno}}]{2012ApJ...757...83D}
{Desai}, S., {Armstrong}, R., {Mohr}, J.~J., {et~al.} 2012, \apj, 757, 83

\bibitem[{{Dias} {et~al.}(2010){Dias}, {Coelho}, {Barbuy}, {Kerber}, \&
  {Idiart}}]{Dias2010}
{Dias}, B., {Coelho}, P., {Barbuy}, B., {Kerber}, L., \& {Idiart}, T. 2010,
  \aap, 520, A85

\bibitem[{{D'Onghia} \& {Lake}(2008)}]{D'Onghia:2008a}
{D'Onghia}, E., \& {Lake}, G. 2008, \apjl, 686, L61

\bibitem[{{Dooley} {et~al.}(2017){Dooley}, {Peter}, {Carlin}, {Frebel},
  {Bechtol}, \& {Willman}}]{DooleyPaper}
{Dooley}, G.~A., {Peter}, A. H.~G., {Carlin}, J.~L., {et~al.} 2017, \mnras,
  472, 1060

\bibitem[{{Drlica-Wagner} {et~al.}(2016){Drlica-Wagner}, {Bechtol}, {Allam},
  {et~al.}}]{Drlica-Wagner:2016}
{Drlica-Wagner}, A., {Bechtol}, K., {Allam}, S., {et~al.} 2016, \apjl, 833, L5

\bibitem[{Drlica-Wagner {et~al.}(2015)}]{Drlica-Wagner:2015}
Drlica-Wagner, A., {et~al.} 2015, \apj, 813, 109

\bibitem[{{Drlica-Wagner} {et~al.}(2020){Drlica-Wagner}, {Bechtol}, {Mau},
  {McNanna}, {Nadler}, {Pace}, {Li}, {Pieres}, {Rozo}, {Simon}, {Walker},
  {Wechsler}, {Abbott}, {Allam}, {Annis}, {Bertin}, {Brooks}, {Burke},
  {Rosell}, {Carrasco Kind}, {Carretero}, {Costanzi}, {da Costa}, {De Vicente},
  {Desai}, {Diehl}, {Doel}, {Eifler}, {Everett}, {Flaugher}, {Frieman},
  {Garc{\'\i}a-Bellido}, {Gaztanaga}, {Gruen}, {Gruendl}, {Gschwend},
  {Gutierrez}, {Honscheid}, {James}, {Krause}, {Kuehn}, {Kuropatkin}, {Lahav},
  {Maia}, {Marshall}, {Melchior}, {Menanteau}, {Miquel}, {Palmese}, {Plazas},
  {Sanchez}, {Scarpine}, {Schubnell}, {Serrano}, {Sevilla-Noarbe}, {Smith},
  {Suchyta}, {Tarle}, \& {DES Collaboration}}]{PaperI}
{Drlica-Wagner}, A., {Bechtol}, K., {Mau}, S., {et~al.} 2020, \apj, 893, 47

\bibitem[{Erkal \& Belokurov(2019)}]{Erkal:2019b}
Erkal, D., \& Belokurov, V.~A. 2019, arXiv e-prints, arXiv:1907.09484

\bibitem[{Erkal \& Belokurov(2020)}]{EB2020_LMCmass}
---. 2020, Monthly Notices of the Royal Astronomical Society, 495, 2554.
\newblock \url{https://doi.org/10.1093/mnras/staa1238}

\bibitem[{{Erkal} {et~al.}(2019){Erkal}, {Belokurov}, {Laporte}, {Koposov},
  {Li}, {Grillmair}, {Kallivayalil}, {Price-Whelan}, {Evans}, {Hawkins},
  {Hendel}, {Mateu}, {Navarro}, {del Pino}, {Slater}, {Sohn}, \& {Orphan Aspen
  Treasury Collaboration}}]{Erkal:2019a}
{Erkal}, D., {Belokurov}, V., {Laporte}, C.~F.~P., {et~al.} 2019, \mnras, 487,
  2685

\bibitem[{{Fadely} {et~al.}(2011){Fadely}, {Willman}, {Geha}, {Walsh},
  {Mu{\~n}oz}, {Jerjen}, {Vargas}, \& {Da Costa}}]{Fadely:2011}
{Fadely}, R., {Willman}, B., {Geha}, M., {et~al.} 2011, \aj, 142, 88

\bibitem[{{Feroz} \& {Hobson}(2008)}]{Feroz2008MNRAS.384..449F}
{Feroz}, F., \& {Hobson}, M.~P. 2008, \mnras, 384, 449

\bibitem[{{Feroz} {et~al.}(2009){Feroz}, {Hobson}, \&
  {Bridges}}]{Feroz2009MNRAS.398.1601F}
{Feroz}, F., {Hobson}, M.~P., \& {Bridges}, M. 2009, \mnras, 398, 1601

\bibitem[{{Flaugher} {et~al.}(2015){Flaugher}, {Diehl}, {Honscheid}, {Abbott},
  {Alvarez}, {Angstadt}, {Annis}, {Antonik}, {Ballester}, {Beaufore},
  {Bernstein}, {Bernstein}, {Bigelow}, {Bonati}, {Boprie}, {Brooks},
  {Buckley-Geer}, {Campa}, {Cardiel-Sas}, {Castand er}, {Castilla}, {Cease},
  {Cela-Ruiz}, {Chappa}, {Chi}, {Cooper}, {da Costa}, {Dede}, {Derylo},
  {DePoy}, {de Vicente}, {Doel}, {Drlica-Wagner}, {Eiting}, {Elliott}, {Emes},
  {Estrada}, {Fausti Neto}, {Finley}, {Flores}, {Frieman}, {Gerdes},
  {Gladders}, {Gregory}, {Gutierrez}, {Hao}, {Holland}, {Holm}, {Huffman},
  {Jackson}, {James}, {Jonas}, {Karcher}, {Karliner}, {Kent}, {Kessler},
  {Kozlovsky}, {Kron}, {Kubik}, {Kuehn}, {Kuhlmann}, {Kuk}, {Lahav}, {Lathrop},
  {Lee}, {Levi}, {Lewis}, {Li}, {Mand richenko}, {Marshall}, {Martinez},
  {Merritt}, {Miquel}, {Mu{\~n}oz}, {Neilsen}, {Nichol}, {Nord}, {Ogando},
  {Olsen}, {Palaio}, {Patton}, {Peoples}, {Plazas}, {Rauch}, {Reil}, {Rheault},
  {Roe}, {Rogers}, {Roodman}, {Sanchez}, {Scarpine}, {Schindler}, {Schmidt},
  {Schmitt}, {Schubnell}, {Schultz}, {Schurter}, {Scott}, {Serrano}, {Shaw},
  {Smith}, {Soares-Santos}, {Stefanik}, {Stuermer}, {Suchyta}, {Sypniewski},
  {Tarle}, {Thaler}, {Tighe}, {Tran}, {Tucker}, {Walker}, {Wang}, {Watson},
  {Weaverdyck}, {Wester}, {Woods}, {Yanny}, \& {DES
  Collaboration}}]{Flaugher:2015}
{Flaugher}, B., {Diehl}, H.~T., {Honscheid}, K., {et~al.} 2015, \aj, 150, 150

\bibitem[{{Foreman-Mackey} {et~al.}(2013){Foreman-Mackey}, {Hogg}, {Lang}, \&
  {Goodman}}]{Foreman-Mackey:2013}
{Foreman-Mackey}, D., {Hogg}, D.~W., {Lang}, D., \& {Goodman}, J. 2013, \pasp,
  125, 306

\bibitem[{{Gaia Collaboration} {et~al.}(2018){Gaia Collaboration}, {Brown},
  {Vallenari}, {Prusti}, {de Bruijne}, {Babusiaux}, {Bailer-Jones}, {Biermann},
  {Evans}, {Eyer}, {Jansen}, {Jordi}, {Klioner}, {Lammers}, {Lindegren},
  {Luri}, {Mignard}, {Panem}, {Pourbaix}, {Randich}, {Sartoretti}, {Siddiqui},
  {Soubiran}, {van Leeuwen}, {Walton}, {Arenou}, {Bastian}, {Cropper},
  {Drimmel}, {Katz}, {Lattanzi}, {Bakker}, {Cacciari}, {Casta{\~n}eda},
  {Chaoul}, {Cheek}, {De Angeli}, {Fabricius}, {Guerra}, {Holl}, {Masana},
  {Messineo}, {Mowlavi}, {Nienartowicz}, {Panuzzo}, {Portell}, {Riello},
  {Seabroke}, {Tanga}, {Th{\'e}venin}, {Gracia-Abril}, {Comoretto},
  {Garcia-Reinaldos}, {Teyssier}, {Altmann}, {Andrae}, {Audard},
  {Bellas-Velidis}, {Benson}, {Berthier}, {Blomme}, {Burgess}, {Busso},
  {Carry}, {Cellino}, {Clementini}, {Clotet}, {Creevey}, {Davidson}, {De
  Ridder}, {Delchambre}, {Dell'Oro}, {Ducourant}, {Fern{\'a}ndez-
  Hern{\'a}ndez}, {Fouesneau}, {Fr{\'e}mat}, {Galluccio}, {Garc{\'\i}a-Torres},
  {Gonz{\'a}lez-N{\'u}{\~n}ez}, {Gonz{\'a}lez-Vidal}, {Gosset}, {Guy},
  {Halbwachs}, {Hambly}, {Harrison}, {Hern{\'a}ndez}, {Hestroffer}, {Hodgkin},
  {Hutton}, {Jasniewicz}, {Jean-Antoine-Piccolo}, {Jordan}, {Korn},
  {Krone-Martins}, {Lanzafame}, {Lebzelter}, {L{\"o}ffler}, {Manteiga},
  {Marrese}, {Mart{\'\i}n-Fleitas}, {Moitinho}, {Mora}, {Muinonen}, {Osinde},
  {Pancino}, {Pauwels}, {Petit}, {Recio-Blanco}, {Richards}, {Rimoldini},
  {Robin}, {Sarro}, {Siopis}, {Smith}, {Sozzetti}, {S{\"u}veges}, {Torra}, {van
  Reeven}, {Abbas}, {Abreu Aramburu}, {Accart}, {Aerts}, {Altavilla},
  {{\'A}lvarez}, {Alvarez}, {Alves}, {Anderson}, {Andrei}, {Anglada Varela},
  {Antiche}, {Antoja}, {Arcay}, {Astraatmadja}, {Bach}, {Baker},
  {Balaguer-N{\'u}{\~n}ez}, {Balm}, {Barache}, {Barata}, {Barbato}, {Barblan},
  {Barklem}, {Barrado}, {Barros}, {Barstow}, {Bartholom{\'e} Mu{\~n}oz},
  {Bassilana}, {Becciani}, {Bellazzini}, {Berihuete}, {Bertone}, {Bianchi},
  {Bienaym{\'e}}, {Blanco-Cuaresma}, {Boch}, {Boeche}, {Bombrun}, {Borrachero},
  {Bossini}, {Bouquillon}, {Bourda}, {Bragaglia}, {Bramante}, {Breddels},
  {Bressan}, {Brouillet}, {Br{\"u}semeister}, {Brugaletta}, {Bucciarelli},
  {Burlacu}, {Busonero}, {Butkevich}, {Buzzi}, {Caffau}, {Cancelliere},
  {Cannizzaro}, {Cantat-Gaudin}, {Carballo}, {Carlucci}, {Carrasco},
  {Casamiquela}, {Castellani}, {Castro-Ginard}, {Charlot}, {Chemin},
  {Chiavassa}, {Cocozza}, {Costigan}, {Cowell}, {Crifo}, {Crosta}, {Crowley},
  {Cuypers}, {Dafonte}, {Damerdji}, {Dapergolas}, {David}, {David}, {de
  Laverny}, {De Luise}, {De March}, {de Martino}, {de Souza}, {de Torres},
  {Debosscher}, {del Pozo}, {Delbo}, {Delgado}, {Delgado}, {Di Matteo},
  {Diakite}, {Diener}, {Distefano}, {Dolding}, {Drazinos}, {Dur{\'a}n},
  {Edvardsson}, {Enke}, {Eriksson}, {Esquej}, {Eynard Bontemps}, {Fabre},
  {Fabrizio}, {Faigler}, {Falc{\~a}o}, {Farr{\`a}s Casas}, {Federici},
  {Fedorets}, {Fernique}, {Figueras}, {Filippi}, {Findeisen}, {Fonti},
  {Fraile}, {Fraser}, {Fr{\'e}zouls}, {Gai}, {Galleti}, {Garabato},
  {Garc{\'\i}a-Sedano}, {Garofalo}, {Garralda}, {Gavel}, {Gavras}, {Gerssen},
  {Geyer}, {Giacobbe}, {Gilmore}, {Girona}, {Giuffrida}, {Glass}, {Gomes},
  {Granvik}, {Gueguen}, {Guerrier}, {Guiraud}, {Guti{\'e}rrez-S{\'a}nchez},
  {Haigron}, {Hatzidimitriou}, {Hauser}, {Haywood}, {Heiter}, {Helmi}, {Heu},
  {Hilger}, {Hobbs}, {Hofmann}, {Holland}, {Huckle}, {Hypki}, {Icardi},
  {Jan{\ss}en}, {Jevardat de Fombelle}, {Jonker}, {Juh{\'a}sz}, {Julbe},
  {Karampelas}, {Kewley}, {Klar}, {Kochoska}, {Kohley}, {Kolenberg},
  {Kontizas}, {Kontizas}, {Koposov}, {Kordopatis}, {Kostrzewa-Rutkowska},
  {Koubsky}, {Lambert}, {Lanza}, {Lasne}, {Lavigne}, {Le Fustec}, {Le
  Poncin-Lafitte}, {Lebreton}, {Leccia}, {Leclerc}, {Lecoeur-Taibi},
  {Lenhardt}, {Leroux}, {Liao}, {Licata}, {Lindstr{\o}m}, {Lister}, {Livanou},
  {Lobel}, {L{\'o}pez}, {Managau}, {Mann}, {Mantelet}, {Marchal}, {Marchant},
  {Marconi}, {Marinoni}, {Marschalk{\'o}}, {Marshall}, {Martino}, {Marton},
  {Mary}, {Massari}, {Matijevi{\v{c}}}, {Mazeh}, {McMillan}, {Messina},
  {Michalik}, {Millar}, {Molina}, {Molinaro}, {Moln{\'a}r}, {Montegriffo},
  {Mor}, {Morbidelli}, {Morel}, {Morris}, {Mulone}, {Muraveva}, {Musella},
  {Nelemans}, {Nicastro}, {Noval}, {O'Mullane}, {Ord{\'e}novic},
  {Ord{\'o}{\~n}ez-Blanco}, {Osborne}, {Pagani}, {Pagano}, {Pailler},
  {Palacin}, {Palaversa}, {Panahi}, {Pawlak}, {Piersimoni}, {Pineau}, {Plachy},
  {Plum}, {Poggio}, {Poujoulet}, {Pr{\v{s}}a}, {Pulone}, {Racero}, {Ragaini},
  {Rambaux}, {Ramos-Lerate}, {Regibo}, {Reyl{\'e}}, {Riclet}, {Ripepi}, {Riva},
  {Rivard}, {Rixon}, {Roegiers}, {Roelens}, {Romero-G{\'o}mez}, {Rowell},
  {Royer}, {Ruiz-Dern}, {Sadowski}, {Sagrist{\`a} Sell{\'e}s}, {Sahlmann},
  {Salgado}, {Salguero}, {Sanna}, {Santana- Ros}, {Sarasso}, {Savietto},
  {Schultheis}, {Sciacca}, {Segol}, {Segovia}, {S{\'e}gransan}, {Shih},
  {Siltala}, {Silva}, {Smart}, {Smith}, {Solano}, {Solitro}, {Sordo}, {Soria
  Nieto}, {Souchay}, {Spagna}, {Spoto}, {Stampa}, {Steele},
  {Steidelm{\"u}ller}, {Stephenson}, {Stoev}, {Suess}, {Surdej}, {Szabados},
  {Szegedi-Elek}, {Tapiador}, {Taris}, {Tauran}, {Taylor}, {Teixeira},
  {Terrett}, {Teyssandier}, {Thuillot}, {Titarenko}, {Torra Clotet}, {Turon},
  {Ulla}, {Utrilla}, {Uzzi}, {Vaillant}, {Valentini}, {Valette}, {van Elteren},
  {Van Hemelryck}, {van Leeuwen}, {Vaschetto}, {Vecchiato}, {Veljanoski},
  {Viala}, {Vicente}, {Vogt}, {von Essen}, {Voss}, {Votruba}, {Voutsinas},
  {Walmsley}, {Weiler}, {Wertz}, {Wevers}, {Wyrzykowski}, {Yoldas},
  {{\v{Z}}erjal}, {Ziaeepour}, {Zorec}, {Zschocke}, {Zucker}, {Zurbach}, \&
  {Zwitter}}]{Gaia:2018}
{Gaia Collaboration}, {Brown}, A.~G.~A., {Vallenari}, A., {et~al.} 2018, \aap,
  616, A1

\bibitem[{{Gatto} {et~al.}(2020){Gatto}, {Ripepi}, {Bellazzini}, {Cignoni},
  {Cioni}, {Dall'Ora}, {Longo}, {Marconi}, {Schipani}, \&
  {Tosi}}]{2020arXiv200700341G}
{Gatto}, M., {Ripepi}, V., {Bellazzini}, M., {et~al.} 2020, arXiv e-prints,
  arXiv:2007.00341

\bibitem[{{Geha} {et~al.}(2009){Geha}, {Willman}, {Simon}, {Strigari}, {Kirby},
  {Law}, \& {Strader}}]{Seg1}
{Geha}, M., {Willman}, B., {Simon}, J.~D., {et~al.} 2009, \apj, 692, 1464

\bibitem[{Gieles \& Renaud(2016)}]{10.1093/mnrasl/slw163}
Gieles, M., \& Renaud, F. 2016, Monthly Notices of the Royal Astronomical
  Society: Letters, 463, L103.
\newblock \url{https://doi.org/10.1093/mnrasl/slw163}

\bibitem[{{Girardi} {et~al.}(1995){Girardi}, {Chiosi}, {Bertelli}, \&
  {Bressan}}]{1995A&A...298...87G}
{Girardi}, L., {Chiosi}, C., {Bertelli}, G., \& {Bressan}, A. 1995, \aap, 298,
  87

\bibitem[{{Glatt} {et~al.}(2008){Glatt}, {Gallagher}, {Grebel}, {Nota},
  {Sabbi}, {Sirianni}, {Clementini}, {Tosi}, {Harbeck}, {Koch}, \&
  {Cracraft}}]{121age}
{Glatt}, K., {Gallagher}, John~S., I., {Grebel}, E.~K., {et~al.} 2008, \aj,
  135, 1106

\bibitem[{{G{\'o}rski} {et~al.}(2005){G{\'o}rski}, {Hivon}, {Banday},
  {Wandelt}, {Hansen}, {Reinecke}, \& {Bartelmann}}]{Gorski:2005}
{G{\'o}rski}, K.~M., {Hivon}, E., {Banday}, A.~J., {et~al.} 2005, \apj, 622,
  759

\bibitem[{{Gravity Collaboration} {et~al.}(2019){Gravity Collaboration},
  {Abuter}, {Amorim}, {Baub{\"o}ck}, {Berger}, {Bonnet}, {Brand ner},
  {Cl{\'e}net}, {Coud{\'e} Du Foresto}, {de Zeeuw}, {Dexter}, {Duvert},
  {Eckart}, {Eisenhauer}, {F{\"o}rster Schreiber}, {Garcia}, {Gao}, {Gendron},
  {Genzel}, {Gerhard}, {Gillessen}, {Habibi}, {Haubois}, {Henning}, {Hippler},
  {Horrobin}, {Jim{\'e}nez-Rosales}, {Jocou}, {Kervella}, {Lacour},
  {Lapeyr{\`e}re}, {Le Bouquin}, {L{\'e}na}, {Ott}, {Paumard}, {Perraut},
  {Perrin}, {Pfuhl}, {Rabien}, {Rodriguez Coira}, {Rousset}, {Scheithauer},
  {Sternberg}, {Straub}, {Straubmeier}, {Sturm}, {Tacconi}, {Vincent}, {von
  Fellenberg}, {Waisberg}, {Widmann}, {Wieprecht}, {Wiezorrek}, {Woillez}, \&
  {Yazici}}]{Abuter:2019}
{Gravity Collaboration}, {Abuter}, R., {Amorim}, A., {et~al.} 2019, \aap, 625,
  L10

\bibitem[{{Harris} \& {Zaritsky}(2004)}]{HZ04}
{Harris}, J., \& {Zaritsky}, D. 2004, \aj, 127, 1531

\bibitem[{{Harris} \& {Zaritsky}(2009)}]{2009AJ....138.1243H}
---. 2009, \aj, 138, 1243

\bibitem[{{Harris}(1996, 2010 edition)}]{Harris:1996}
{Harris}, W.~E. 1996, 2010 edition, \aj, 112, 1487

\bibitem[{{Hernquist}(1990)}]{Hernquist:1990}
{Hernquist}, L. 1990, \apj, 356, 359

\bibitem[{{Homma} {et~al.}(2016){Homma}, {Chiba}, {Okamoto}, {Komiyama},
  {Tanaka}, {Tanaka}, {Ishigaki}, {Akiyama}, {Arimoto}, {Garmilla}, {Lupton},
  {Strauss}, {Furusawa}, {Miyazaki}, {Murayama}, {Nishizawa}, {Takada},
  {Usuda}, \& {Wang}}]{Homma:2016}
{Homma}, D., {Chiba}, M., {Okamoto}, S., {et~al.} 2016, \apj, 832, 21

\bibitem[{{Homma} {et~al.}(2018){Homma}, {Chiba}, {Okamoto}, {Komiyama},
  {Tanaka}, {Tanaka}, {Ishigaki}, {Hayashi}, {Arimoto}, {Garmilla}, {Lupton},
  {Strauss}, {Miyazaki}, {Wang}, \& {Murayama}}]{Homma:2017}
---. 2018, \pasj, 70, S18

\bibitem[{{Homma} {et~al.}(2019){Homma}, {Chiba}, {Komiyama}, {Tanaka},
  {Okamoto}, {Tanaka}, {Ishigaki}, {Hayashi}, {Arimoto}, {Carlsten}, {Lupton},
  {Strauss}, {Miyazaki}, {Torrealba}, {Wang}, \& {Murayama}}]{Homma:2019}
{Homma}, D., {Chiba}, M., {Komiyama}, Y., {et~al.} 2019, \pasj, 91

\bibitem[{Hunter(2007)}]{Hunter:2007}
Hunter, J.~D. 2007, Computing In Science \& Engineering, 9, 90

\bibitem[{{Innanen} {et~al.}(1983){Innanen}, {Harris}, \&
  {Webbink}}]{1983_equation}
{Innanen}, K.~A., {Harris}, W.~E., \& {Webbink}, R.~F. 1983, \aj, 88, 338

\bibitem[{Jahn {et~al.}(2019)Jahn, Sales, Wetzel, Boylan-Kolchin, Chan,
  El-Badry, Lazar, \& Bullock}]{Jahn:2019}
Jahn, E.~D., Sales, L.~V., Wetzel, A., {et~al.} 2019, Monthly Notices of the
  Royal Astronomical Society, 489, 5348–5364.
\newblock \url{http://dx.doi.org/10.1093/mnras/stz2457}

\bibitem[{{Jethwa} {et~al.}(2016){Jethwa}, {Erkal}, \&
  {Belokurov}}]{Jethwa:2016}
{Jethwa}, P., {Erkal}, D., \& {Belokurov}, V. 2016, \mnras, 461, 2212

\bibitem[{{Jethwa} {et~al.}(2018){Jethwa}, {Erkal}, \&
  {Belokurov}}]{Jethwa:2018}
---. 2018, \mnras, 473, 2060

\bibitem[{Jones {et~al.}(2001)Jones, Oliphant, Peterson, {et~al.}}]{scipy:2001}
Jones, E., Oliphant, T., Peterson, P., {et~al.} 2001, {SciPy}: Open source
  scientific tools for {Python}, , .
\newblock \url{http://www.scipy.org/}

\bibitem[{{Kallivayalil} {et~al.}(2013){Kallivayalil}, {van der Marel},
  {Besla}, {Anderson}, \& {Alcock}}]{lmcpm}
{Kallivayalil}, N., {van der Marel}, R.~P., {Besla}, G., {Anderson}, J., \&
  {Alcock}, C. 2013, \apj, 764, 161

\bibitem[{Kallivayalil {et~al.}(2018)Kallivayalil, Sales, Zivick, Fritz,
  Del~Pino, Sohn, Besla, van~der Marel, Navarro, \& Sacchi}]{Kallivayalil:2018}
Kallivayalil, N., Sales, L.~V., Zivick, P., {et~al.} 2018, The Astrophysical
  Journal, 867, 19.
\newblock \url{http://dx.doi.org/10.3847/1538-4357/aadfee}

\bibitem[{{Kim} \& {Jerjen}(2015{\natexlab{a}})}]{Kim:2015c}
{Kim}, D., \& {Jerjen}, H. 2015{\natexlab{a}}, \apjl, 808, L39

\bibitem[{{Kim} \& {Jerjen}(2015{\natexlab{b}})}]{Kim:2015a}
---. 2015{\natexlab{b}}, \apj, 799, 73

\bibitem[{{Kim} {et~al.}(2016){Kim}, {Jerjen}, {Mackey}, {Da Costa}, \&
  {Milone}}]{Kim:2016}
{Kim}, D., {Jerjen}, H., {Mackey}, D., {Da Costa}, G.~S., \& {Milone}, A.~P.
  2016, \apj, 820, 119

\bibitem[{{Kirby} {et~al.}(2013){Kirby}, {Boylan-Kolchin}, {Cohen}, {Geha},
  {Bullock}, \& {Kaplinghat}}]{SegII}
{Kirby}, E.~N., {Boylan-Kolchin}, M., {Cohen}, J.~G., {et~al.} 2013, \apj, 770,
  16

\bibitem[{{Koposov} {et~al.}(2017){Koposov}, {Belokurov}, \&
  {Torrealba}}]{Koposov:2017}
{Koposov}, S.~E., {Belokurov}, V., \& {Torrealba}, G. 2017, \mnras, 470, 2702

\bibitem[{{Koposov} {et~al.}(2015{\natexlab{a}}){Koposov}, {Belokurov},
  {Torrealba}, \& {Evans}}]{Koposov:2015cua}
{Koposov}, S.~E., {Belokurov}, V., {Torrealba}, G., \& {Evans}, N.~W.
  2015{\natexlab{a}}, \apj, 805, 130

\bibitem[{{Koposov} {et~al.}(2015{\natexlab{b}}){Koposov}, {Casey},
  {Belokurov}, {Lewis}, {Gilmore}, {Worley}, {Hourihane}, {Bensby},
  {Bragaglia}, {Bergemann}, {Carraro}, {Flaccomio}, {Heiter}, {Hill}, {Jofre},
  {de Laverny}, {Monaco}, {Sbordone}, {Mikolaitis}, \& {Ryde}}]{Koposov:2015b}
{Koposov}, S.~E., {Casey}, A.~R., {Belokurov}, V., {et~al.} 2015{\natexlab{b}},
  submitted to \apj, arXiv:1504.07916

\bibitem[{{Koposov} {et~al.}(2018){Koposov}, {Walker}, {Belokurov}, {Casey},
  {Geringer-Sameth}, {Mackey}, {Da Costa}, {Erkal}, {Jethwa}, {Mateo},
  {Olszewski}, \& {Bailey}}]{Koposov:2018}
{Koposov}, S.~E., {Walker}, M.~G., {Belokurov}, V., {et~al.} 2018, \mnras, 479,
  5343

\bibitem[{{Kruijssen}(2008)}]{GCML}
{Kruijssen}, J.~M.~D. 2008, \aap, 486, L21

\bibitem[{{Laevens} {et~al.}(2014){Laevens}, {Martin}, {Sesar}, {Bernard},
  {Rix}, {Slater}, {Bell}, {Ferguson}, {Schlafly}, {Burgett}, {Chambers},
  {Denneau}, {Draper}, {Kaiser}, {Kudritzki}, {Magnier}, {Metcalfe}, {Morgan},
  {Price}, {Sweeney}, {Tonry}, {Wainscoat}, \& {Waters}}]{Laevens:2014}
{Laevens}, B.~P.~M., {Martin}, N.~F., {Sesar}, B., {et~al.} 2014, \apjl, 786,
  L3

\bibitem[{{Laevens} {et~al.}(2015{\natexlab{a}}){Laevens}, {Martin}, {Ibata},
  {Rix}, {Bernard}, {Bell}, {Sesar}, {Ferguson}, {Schlafly}, {Slater},
  {Burgett}, {Chambers}, {Flewelling}, {Hodapp}, {Kaiser}, {Kudritzki},
  {Lupton}, {Magnier}, {Metcalfe}, {Morgan}, {Price}, {Tonry}, {Wainscoat}, \&
  {Waters}}]{Laevens:2015a}
{Laevens}, B.~P.~M., {Martin}, N.~F., {Ibata}, R.~A., {et~al.}
  2015{\natexlab{a}}, \apjl, 802, L18

\bibitem[{{Laevens} {et~al.}(2015{\natexlab{b}}){Laevens}, {Martin}, {Bernard},
  {Schlafly}, {Sesar}, {Rix}, {Bell}, {Ferguson}, {Slater}, {Sweeney}, {Wyse},
  {Huxor}, {Burgett}, {Chambers}, {Draper}, {Hodapp}, {Kaiser}, {Magnier},
  {Metcalfe}, {Tonry}, {Wainscoat}, \& {Waters}}]{Laevens:2015b}
{Laevens}, B. P.~M., {Martin}, N.~F., {Bernard}, E.~J., {et~al.}
  2015{\natexlab{b}}, \apj, 813, 44

\bibitem[{{Luque} {et~al.}(2018){Luque}, {Santiago}, {Pieres}, {Marshall},
  {et~al.}}]{Luque:2018}
{Luque}, E., {Santiago}, B., {Pieres}, A., {Marshall}, J.~L., {et~al.} 2018,
  \mnras, 478, 2006

\bibitem[{{Luque} {et~al.}(2016){Luque}, {Queiroz}, {Santiago}, {Pieres},
  {Balbinot}, {Bechtol}, {Drlica-Wagner}, {Neto}, {da Costa}, {Maia}, {Yanny},
  {Abbott}, {Allam}, {Benoit-L{\'e}vy}, {Bertin}, {Brooks}, {Buckley-Geer},
  {Burke}, {Rosell}, {Kind}, {Carretero}, {Cunha}, {Desai}, {Diehl},
  {Dietrich}, {Eifler}, {Finley}, {Flaugher}, {Fosalba}, {Frieman}, {Gerdes},
  {Gruen}, {Gutierrez}, {Honscheid}, {James}, {Kuehn}, {Kuropatkin}, {Lahav},
  {Li}, {March}, {Marshall}, {Martini}, {Miquel}, {Neilsen}, {Nichol}, {Nord},
  {Ogando}, {Plazas}, {Romer}, {Roodman}, {Sanchez}, {Scarpine}, {Schubnell},
  {Sevilla-Noarbe}, {Smith}, {Soares-Santos}, {Sobreira}, {Suchyta}, {Swanson},
  {Tarle}, {Thaler}, {Tucker}, {Walker}, \& {Zhang}}]{Luque:2016}
{Luque}, E., {Queiroz}, A., {Santiago}, B., {et~al.} 2016, \mnras, 458, 603

\bibitem[{Luque {et~al.}(2017{\natexlab{a}})Luque, Pieres, Santiago, Yanny,
  Vivas, Queiroz, Drlica-Wagner, Morganson, Balbinot, Marshall, Li, Neto,
  da~Costa, Maia, Bechtol, Kim, Bernstein, Dodelson, Whiteway, Diehl, Finley,
  Abbott, Abdalla, Allam, Annis, Benoit-Lévy, Bertin, Brooks, Burke, Rosell,
  Kind, Carretero, Cunha, D'Andrea, Desai, Doel, Evrard, Flaugher, Fosalba,
  Gerdes, Goldstein, Gruen, Gruendl, Gutierrez, James, Kuehn, Kuropatkin,
  Lahav, Martini, Miquel, Nord, Ogando, Plazas, Romer, Sanchez, Scarpine,
  Schubnell, Sevilla-Noarbe, Smith, Soares-Santos, Sobreira, Suchyta, Swanson,
  Tarle, Thomas, \& Walker}]{Luque:2017}
Luque, E., Pieres, A., Santiago, B., {et~al.} 2017{\natexlab{a}}, \mnras, 468,
  97.
\newblock \url{https://doi.org/10.1093/mnras/stx405}

\bibitem[{Luque {et~al.}(2017{\natexlab{b}})Luque, Pieres, Santiago, Yanny,
  Vivas, Queiroz, Drlica-Wagner, Morganson, Balbinot, Marshall, Li, Neto,
  da~Costa, Maia, Bechtol, Kim, Bernstein, Dodelson, Whiteway, Diehl, Finley,
  Abbott, Abdalla, Allam, Annis, Benoit-Lévy, Bertin, Brooks, Burke, Rosell,
  Kind, Carretero, Cunha, D'Andrea, Desai, Doel, Evrard, Flaugher, Fosalba,
  Gerdes, Goldstein, Gruen, Gruendl, Gutierrez, James, Kuehn, Kuropatkin,
  Lahav, Martini, Miquel, Nord, Ogando, Plazas, Romer, Sanchez, Scarpine,
  Schubnell, Sevilla-Noarbe, Smith, Soares-Santos, Sobreira, Suchyta, Swanson,
  Tarle, Thomas, \& Walker}]{DES2}
---. 2017{\natexlab{b}}, Monthly Notices of the Royal Astronomical Society,
  468, 97.
\newblock \url{https://doi.org/10.1093/mnras/stx405}

\bibitem[{{Lynden-Bell}(1976)}]{DLB76}
{Lynden-Bell}, D. 1976, \mnras, 174, 695

\bibitem[{{Martin} {et~al.}(2008){Martin}, {de Jong}, \& {Rix}}]{Martin:2008}
{Martin}, N.~F., {de Jong}, J.~T.~A., \& {Rix}, H.-W. 2008, \apj, 684, 1075

\bibitem[{{Martin} {et~al.}(2015){Martin}, {Nidever}, {Besla}, {Olsen},
  {Walker}, {Vivas}, {Gruendl}, {Kaleida}, {Mu{\~n}oz}, {Blum}, {Saha}, {Conn},
  {Bell}, {Chu}, {Cioni}, {de Boer}, {Gallart}, {Jin}, {Kunder}, {Majewski},
  {Martinez-Delgado}, {Monachesi}, {Monelli}, {Monteagudo}, {No{\"e}l},
  {Olszewski}, {Stringfellow}, {van der Marel}, \&
  {Zaritsky}}]{martin_2015_hydra_ii}
{Martin}, N.~F., {Nidever}, D.~L., {Besla}, G., {et~al.} 2015, \apjl, 804, L5

\bibitem[{{Martin} {et~al.}(2016){Martin}, {Jungbluth}, {Nidever}, {Bell},
  {Besla}, {Blum}, {Cioni}, {Conn}, {Kaleida}, {Gallart}, {Jin}, {Majewski},
  {Martinez-Delgado}, {Monachesi}, {Mu{\~n}oz}, {No{\"e}l}, {Olsen},
  {Stringfellow}, {van der Marel}, {Vivas}, {Walker}, \& {Zaritsky}}]{SMASH1}
{Martin}, N.~F., {Jungbluth}, V., {Nidever}, D.~L., {et~al.} 2016, \apjl, 830,
  L10

\bibitem[{Martínez-Vázquez {et~al.}(2019)Martínez-Vázquez, Vivas, Gurevich,
  Walker, McCarthy, Pace, Stringer, Santiago, Hounsell, Macri, \&
  et~al.}]{Martinez-Vazquez:2019}
Martínez-Vázquez, C.~E., Vivas, A.~K., Gurevich, M., {et~al.} 2019, Monthly
  Notices of the Royal Astronomical Society, doi:10.1093/mnras/stz2609.
\newblock \url{http://dx.doi.org/10.1093/mnras/stz2609}

\bibitem[{{Massana} {et~al.}(2020){Massana}, {No{\"e}l}, {Nidever}, {Erkal},
  {de Boer}, {Choi}, {Majewski}, {Olsen}, {Monachesi}, {Gallart}, {van der
  Marel}, {Ruiz-Lara}, {Zaritsky}, {Martin}, {Mu{\~n}oz}, {Cioni}, {Bell},
  {Bell}, {Stringfellow}, {Belokurov}, {Monelli}, {Walker},
  {Mart{\'\i}nez-Delgado}, {Vivas}, \& {Conn}}]{2020SMC}
{Massana}, P., {No{\"e}l}, N. E.~D., {Nidever}, D.~L., {et~al.} 2020, arXiv
  e-prints, arXiv:2008.00012

\bibitem[{{Mau} {et~al.}(2019){Mau}, {Drlica-Wagner}, {Bechtol}, {Pace}, {Li},
  {Soares-Santos}, {Kuropatkin}, {Allam}, {Tucker}, {Santana-Silva}, {Yanny},
  {Jethwa}, {Palmese}, {Vivas}, {Burgad}, {Chen}, \& {BLISS
  Collaboration}}]{Mau:2019}
{Mau}, S., {Drlica-Wagner}, A., {Bechtol}, K., {et~al.} 2019, \apj, 875, 154

\bibitem[{{Mau} {et~al.}(2020){Mau}, {Cerny}, {Pace}, {Choi}, {Drlica-Wagner},
  {Santana-Silva}, {Riley}, {Erkal}, {Stringfellow}, {Adam{\'o}w}, {Carlin},
  {Gruendl}, {Hernandez-Lang}, {Kuropatkin}, {Li}, {Mart{\'\i}nez-V{\'a}zquez},
  {Morganson}, {Mutlu-Pakdil}, {Neilsen}, {Nidever}, {Olsen}, {Sand},
  {Tollerud}, {Tucker}, {Yanny}, {Zenteno}, {Allam}, {Barkhouse}, {Bechtol},
  {Bell}, {Balaji}, {Crnojevi{\'c}}, {Esteves}, {Ferguson}, {Gallart},
  {Hughes}, {James}, {Jethwa}, {Johnson}, {Kuehn}, {Majewski}, {Mao},
  {Massana}, {McNanna}, {Monachesi}, {Nadler}, {No{\"e}l}, {Palmese},
  {Paz-Chinchon}, {Pieres}, {Sanchez}, {Shipp}, {Simon}, {Soares-Santos},
  {Tavangar}, {van der Marel}, {Vivas}, {Walker}, \& {Wechsler}}]{MauCerny2020}
{Mau}, S., {Cerny}, W., {Pace}, A.~B., {et~al.} 2020, \apj, 890, 136

\bibitem[{{McConnachie}(2012)}]{McConnachie:2012}
{McConnachie}, A.~W. 2012, \aj, 144, 4

\bibitem[{{Meschin} {et~al.}(2014){Meschin}, {Gallart}, {Aparicio}, {Hidalgo},
  {Monelli}, {Stetson}, \& {Carrera}}]{2014MNRAS.438.1067M}
{Meschin}, I., {Gallart}, C., {Aparicio}, A., {et~al.} 2014, \mnras, 438, 1067

\bibitem[{{Morganson} {et~al.}(2018){Morganson}, {Gruendl}, {Menanteau},
  {Carrasco Kind}, {Chen}, {Daues}, {Drlica-Wagner}, {Friedel}, {Gower},
  {Johnson}, {Johnson}, {Kessler}, {Paz-Chinch{\'o}n}, {Petravick}, {Pond},
  {Yanny}, {Allam}, {Armstrong}, {Barkhouse}, {Bechtol}, {Benoit-L{\'e}vy},
  {Bernstein}, {Bertin}, {Buckley-Geer}, {Covarrubias}, {Desai}, {Diehl},
  {Goldstein}, {Gruen}, {Li}, {Lin}, {Marriner}, {Mohr}, {Neilsen}, {Ngeow},
  {Paech}, {Rykoff}, {Sako}, {Sevilla-Noarbe}, {Sheldon}, {Sobreira}, {Tucker},
  {Wester}, \& {DES Collaboration}}]{Morganson:2018}
{Morganson}, E., {Gruendl}, R.~A., {Menanteau}, F., {et~al.} 2018, \pasp, 130,
  074501

\bibitem[{{Mu{\~n}oz} {et~al.}(2018){Mu{\~n}oz}, {C{\^o}t{\'e}}, {Santana},
  {Geha}, {Simon}, {Oyarz{\'u}n}, {Stetson}, \&
  {Djorgovski}}]{2018ApJ...860...66M}
{Mu{\~n}oz}, R.~R., {C{\^o}t{\'e}}, P., {Santana}, F.~A., {et~al.} 2018, \apj,
  860, 66

\bibitem[{{Mu{\~n}oz} {et~al.}(2012){Mu{\~n}oz}, {Geha}, {C{\^o}t{\'e}},
  {Vargas}, {Santana}, {Stetson}, {Simon}, \& {Djorgovski}}]{Munoz:2012}
{Mu{\~n}oz}, R.~R., {Geha}, M., {C{\^o}t{\'e}}, P., {et~al.} 2012, \apjl, 753,
  L15

\bibitem[{{Mutlu-Pakdil} {et~al.}(2018){Mutlu-Pakdil}, {Sand}, {Carlin},
  {Spekkens}, {Caldwell}, {Crnojevi{\'c}}, {Hughes}, {Willman}, \&
  {Zaritsky}}]{Mutlu-Pakdil:2018}
{Mutlu-Pakdil}, B., {Sand}, D.~J., {Carlin}, J.~L., {et~al.} 2018, \apj, 863,
  25

\bibitem[{{Mutlu-Pakdil} {et~al.}(2019){Mutlu-Pakdil}, {Sand}, {Walker},
  {Caldwell}, {Carlin}, {Collins}, {Crnojevi{\'c}}, {Mateo}, {Olszewski},
  {Seth}, {Strader}, {Willman}, \& {Zaritsky}}]{Mutlu-Pakdil:2019}
{Mutlu-Pakdil}, B., {Sand}, D.~J., {Walker}, M.~G., {et~al.} 2019, \apj, 885,
  53

\bibitem[{{Nadler} {et~al.}(2020){Nadler}, {Wechsler}, {Bechtol}, {Mao},
  {Green}, {Drlica-Wagner}, {McNanna}, {Mau}, {Pace}, {Simon}, {Kravtsov},
  {Dodelson}, {Li}, {Riley}, {Wang}, {Abbott}, {Aguena}, {Allam}, {Annis},
  {Avila}, {Bernstein}, {Bertin}, {Brooks}, {Burke}, {Rosell}, {Kind},
  {Carretero}, {Costanzi}, {da Costa}, {De Vicente}, {Desai}, {Evrard},
  {Flaugher}, {Fosalba}, {Frieman}, {Garc{\'\i}a-Bellido}, {Gaztanaga},
  {Gerdes}, {Gruen}, {Gschwend}, {Gutierrez}, {Hartley}, {Hinton}, {Honscheid},
  {Krause}, {Kuehn}, {Kuropatkin}, {Lahav}, {Maia}, {Marshall}, {Menanteau},
  {Miquel}, {Palmese}, {Paz-Chinch{\'o}n}, {Plazas}, {Romer}, {Sanchez},
  {Santiago}, {Scarpine}, {Serrano}, {Smith}, {Soares-Santos}, {Suchyta},
  {Tarle}, {Thomas}, {Varga}, {Walker}, \& {DES Collaboration}}]{Nadler2020}
{Nadler}, E.~O., {Wechsler}, R.~H., {Bechtol}, K., {et~al.} 2020, \apj, 893, 48

\bibitem[{{Nidever} {et~al.}(2008){Nidever}, {Majewski}, \& {Butler
  Burton}}]{Nidever:2008}
{Nidever}, D.~L., {Majewski}, S.~R., \& {Butler Burton}, W. 2008, \apj, 679,
  432

\bibitem[{{Nidever} {et~al.}(2017){Nidever}, {Olsen}, {Walker}, {Vivas},
  {Blum}, {Kaleida}, {Choi}, {Conn}, {Gruendl}, {Bell}, {Besla}, {Mu{\~n}oz},
  {Gallart}, {Martin}, {Olszewski}, {Saha}, {Monachesi}, {Monelli}, {de Boer},
  {Johnson}, {Zaritsky}, {Stringfellow}, {van der Marel}, {Cioni}, {Jin},
  {Majewski}, {Martinez-Delgado}, {Monteagudo}, {No{\"e}l}, {Bernard},
  {Kunder}, {Chu}, {Bell}, {Santana}, {Frechem}, {Medina}, {Parkash},
  {Ser{\'o}n Navarrete}, \& {Hayes}}]{Nidever2017AJ....154..199N}
{Nidever}, D.~L., {Olsen}, K., {Walker}, A.~R., {et~al.} 2017, \aj, 154, 199

\bibitem[{{Olszewski} {et~al.}(1996){Olszewski}, {Suntzeff}, \&
  {Mateo}}]{1996ARA&A..34..511O}
{Olszewski}, E.~W., {Suntzeff}, N.~B., \& {Mateo}, M. 1996, \araa, 34, 511

\bibitem[{{Pace} \& {Li}(2019)}]{Pace:2019}
{Pace}, A.~B., \& {Li}, T.~S. 2019, \apj, 875, 77

\bibitem[{{Pardy} {et~al.}(2019){Pardy}, {D'Onghia}, {Navarro}, {Grand },
  {Gomez}, {Marinacci}, {Pakmor}, {Simpson}, \& {Springel}}]{Pardy:2019}
{Pardy}, S.~A., {D'Onghia}, E., {Navarro}, J., {et~al.} 2019, arXiv e-prints,
  arXiv:1904.01028

\bibitem[{Patel {et~al.}(2020)Patel, Kallivayalil, Garavito-Camargo, Besla,
  Weisz, van~der Marel, Boylan-Kolchin, Pawlowski, \& G{\'{o}}mez}]{Patel_2020}
Patel, E., Kallivayalil, N., Garavito-Camargo, N., {et~al.} 2020, The
  Astrophysical Journal, 893, 121.
\newblock \url{https://doi.org/10.3847%2F1538-4357%2Fab7b75}

\bibitem[{{Pieres} {et~al.}(2016){Pieres}, {Santiago}, {Balbinot}, {Luque},
  {Queiroz}, {da Costa}, {Maia}, {Drlica-Wagner}, {Roodman}, {Abbott}, {Allam},
  {Benoit-L{\'e}vy}, {Bertin}, {Brooks}, {Buckley-Geer}, {Burke}, {Carnero
  Rosell}, {Carrasco Kind}, {Carretero}, {Cunha}, {Desai}, {Diehl}, {Eifler},
  {Finley}, {Flaugher}, {Fosalba}, {Frieman}, {Gerdes}, {Gruen}, {Gruendl},
  {Gutierrez}, {Honscheid}, {James}, {Kuehn}, {Kuropatkin}, {Lahav}, {Li},
  {Marshall}, {Martini}, {Miller}, {Miquel}, {Nichol}, {Nord}, {Ogando},
  {Plazas}, {Romer}, {Sanchez}, {Scarpine}, {Schubnell}, {Sevilla-Noarbe},
  {Smith}, {Soares-Santos}, {Sobreira}, {Suchyta}, {Swanson}, {Tarle},
  {Thaler}, {Thomas}, {Tucker}, \& {Walker}}]{Pieres2016}
{Pieres}, A., {Santiago}, B., {Balbinot}, E., {et~al.} 2016, \mnras, 461, 519

\bibitem[{{Pietrzy{\'n}ski} {et~al.}(2013){Pietrzy{\'n}ski}, {Graczyk},
  {Gieren}, {Thompson}, {Pilecki}, {Udalski}, {Soszy{\'n}ski}, {Koz{\l}owski},
  {Konorski}, {Suchomska}, {Bono}, {Moroni}, {Villanova}, {Nardetto},
  {Bresolin}, {Kudritzki}, {Storm}, {Gallenne}, {Smolec}, {Minniti}, {Kubiak},
  {Szyma{\'n}ski}, {Poleski}, {Wyrzykowski}, {Ulaczyk}, {Pietrukowicz},
  {G{\'o}rski}, \& {Karczmarek}}]{Pietrzynski:2013}
{Pietrzy{\'n}ski}, G., {Graczyk}, D., {Gieren}, W., {et~al.} 2013, \nat, 495,
  76

\bibitem[{{Plummer}(1911)}]{Plummer:1911}
{Plummer}, H.~C. 1911, \mnras, 71, 460

\bibitem[{{Price-Whelan} {et~al.}(2018){Price-Whelan}, {Sip{\H{o}}cz},
  {G{\"u}nther}, {Lim}, {Crawford}, {Conseil}, {Shupe}, {Craig}, {Dencheva},
  {Ginsburg}, {VanderPlas}, {Bradley}, {P{\'e}rez-Su{\'a}rez}, {de Val-Borro},
  {Paper Contributors}, {Aldcroft}, {Cruz}, {Robitaille}, {Tollerud},
  {Coordination Committee}, {Ardelean}, {Babej}, {Bach}, {Bachetti}, {Bakanov},
  {Bamford}, {Barentsen}, {Barmby}, {Baumbach}, {Berry}, {Biscani}, {Boquien},
  {Bostroem}, {Bouma}, {Brammer}, {Bray}, {Breytenbach}, {Buddelmeijer},
  {Burke}, {Calderone}, {Cano Rodr{\'\i}guez}, {Cara}, {Cardoso}, {Cheedella},
  {Copin}, {Corrales}, {Crichton}, {D{\textquoteright}Avella}, {Deil},
  {Depagne}, {Dietrich}, {Donath}, {Droettboom}, {Earl}, {Erben}, {Fabbro},
  {Ferreira}, {Finethy}, {Fox}, {Garrison}, {Gibbons}, {Goldstein}, {Gommers},
  {Greco}, {Greenfield}, {Groener}, {Grollier}, {Hagen}, {Hirst}, {Homeier},
  {Horton}, {Hosseinzadeh}, {Hu}, {Hunkeler}, {Ivezi{\'c}}, {Jain}, {Jenness},
  {Kanarek}, {Kendrew}, {Kern}, {Kerzendorf}, {Khvalko}, {King}, {Kirkby},
  {Kulkarni}, {Kumar}, {Lee}, {Lenz}, {Littlefair}, {Ma}, {Macleod},
  {Mastropietro}, {McCully}, {Montagnac}, {Morris}, {Mueller}, {Mumford},
  {Muna}, {Murphy}, {Nelson}, {Nguyen}, {Ninan}, {N{\"o}the}, {Ogaz}, {Oh},
  {Parejko}, {Parley}, {Pascual}, {Patil}, {Patil}, {Plunkett}, {Prochaska},
  {Rastogi}, {Reddy Janga}, {Sabater}, {Sakurikar}, {Seifert}, {Sherbert},
  {Sherwood-Taylor}, {Shih}, {Sick}, {Silbiger}, {Singanamalla}, {Singer},
  {Sladen}, {Sooley}, {Sornarajah}, {Streicher}, {Teuben}, {Thomas},
  {Tremblay}, {Turner}, {Terr{\'o}n}, {van Kerkwijk}, {de la Vega}, {Watkins},
  {Weaver}, {Whitmore}, {Woillez}, {Zabalza}, \& {Contributors}}]{astropy:2018}
{Price-Whelan}, A.~M., {Sip{\H{o}}cz}, B.~M., {G{\"u}nther}, H.~M., {et~al.}
  2018, \aj, 156, 123

\bibitem[{{Rubele} {et~al.}(2018){Rubele}, {Pastorelli}, {Girardi}, {Cioni},
  {Zaggia}, {Marigo}, {Bekki}, {Bressan}, {Clementini}, {de Grijs}, {Emerson},
  {Groenewegen}, {Ivanov}, {Muraveva}, {Nanni}, {Oliveira}, {Ripepi}, {Sun}, \&
  {van Loon}}]{LMCSFH_2}
{Rubele}, S., {Pastorelli}, G., {Girardi}, L., {et~al.} 2018, \mnras, 478, 5017

\bibitem[{{Ruiz-Lara} {et~al.}(2020){Ruiz-Lara}, {Gallart}, {Monelli},
  {Nidever}, {Dorta}, {Choi}, {Olsen}, {Besla}, {Bernard}, {Cassisi},
  {Massana}, {No{\"e}l}, {P{\'e}rez}, {Rusakov}, {Cioni}, {Majewski}, {van der
  Marel}, {Mart{\'\i}nez-Delgado}, {Monachesi}, {Monteagudo}, {Mu{\~n}oz},
  {Stringfellow}, {Surot}, {Vivas}, {Walker}, \& {Zaritsky}}]{LMCSFH_3}
{Ruiz-Lara}, T., {Gallart}, C., {Monelli}, M., {et~al.} 2020, \aap, 639, L3

\bibitem[{{Sales} {et~al.}(2017){Sales}, {Navarro}, {Kallivayalil}, \&
  {Frenk}}]{Sales2017MNRAS.465.1879S}
{Sales}, L.~V., {Navarro}, J.~F., {Kallivayalil}, N., \& {Frenk}, C.~S. 2017,
  \mnras, 465, 1879

\bibitem[{Santos {et~al.}(2020)Santos, Maia, Dias, Kerber, Piatti, Bica,
  Angelo, Minniti, Pérez-Villegas, Roman-Lopes, Westera, Fraga, Quint, \&
  Sanmartim}]{10.1093/mnras/staa2425}
Santos, João F~C, J., Maia, F. F.~S., Dias, B., {et~al.} 2020, Monthly Notices
  of the Royal Astronomical Society,
  https://academic.oup.com/mnras/advance-article-pdf/doi/10.1093/mnras/staa2425/33652011/staa2425.pdf,
  staa2425.
\newblock \url{https://doi.org/10.1093/mnras/staa2425}

\bibitem[{{Schlafly} \& {Finkbeiner}(2011)}]{Schlafly:2011}
{Schlafly}, E.~F., \& {Finkbeiner}, D.~P. 2011, \apj, 737, 103

\bibitem[{{Schlegel} {et~al.}(1998){Schlegel}, {Finkbeiner}, \&
  {Davis}}]{Schlegel:1998}
{Schlegel}, D.~J., {Finkbeiner}, D.~P., \& {Davis}, M. 1998, \apj, 500, 525

\bibitem[{{Simon}(2019)}]{Simon:2019}
{Simon}, J.~D. 2019, \araa, 57, 375

\bibitem[{Simon {et~al.}(2017)Simon, Li, Drlica-Wagner, Bechtol, Marshall,
  James, Wang, Strigari, Balbinot, Kuehn, Walker, Abbott, Allam, Annis,
  Benoit-L{\'{e}}vy, Brooks, Buckley-Geer, Burke, Rosell, Kind, Carretero,
  Cunha, D'Andrea, da~Costa, DePoy, Desai, Doel, Fernandez, Flaugher, Frieman,
  Garc{\'{\i}}a-Bellido, Gaztanaga, Goldstein, Gruen, Gutierrez, Kuropatkin,
  Maia, Martini, Menanteau, Miller, Miquel, Neilsen, Nord, Ogando, Plazas,
  Romer, Rykoff, Sanchez, Santiago, Scarpine, Schubnell, Sevilla-Noarbe, Smith,
  Sobreira, Suchyta, Swanson, Tarle, Whiteway, \& and}]{Simon_2017}
Simon, J.~D., Li, T.~S., Drlica-Wagner, A., {et~al.} 2017, The Astrophysical
  Journal, 838, 11.
\newblock \url{https://doi.org/10.38472F1538-43572Faa5be7}

\bibitem[{{Sitek} {et~al.}(2016){Sitek}, {Szyma{\'n}ski}, {Skowron}, {Udalski},
  {Kostrzewa-Rutkowska}, {Skowron}, {Karczmarek}, {Cie{\'s}lar}, {Wyrzykowski},
  {Koz{\l}owski}, {Pietrukowicz}, {Soszy{\'n}ski}, {Mr{\'o}z}, {Pawlak},
  {Poleski}, \& {Ulaczyk}}]{OGLE_LMC}
{Sitek}, M., {Szyma{\'n}ski}, M.~K., {Skowron}, D.~M., {et~al.} 2016, \actaa,
  66, 255

\bibitem[{{Sitek} {et~al.}(2017){Sitek}, {Szyma{\'n}ski}, {Udalski}, {Skowron},
  {Kostrzewa-Rutkowska}, {Skowron}, {Karczmarek}, {Cie{\'s}lar}, {Wyrzykowski},
  {Koz{\l}owski}, {Pietrukowicz}, {Soszy{\'n}ski}, {Mr{\'o}z}, {Pawlak},
  {Poleski}, \& {Ulaczyk}}]{2017AcA....67..363S}
{Sitek}, M., {Szyma{\'n}ski}, M.~K., {Udalski}, A., {et~al.} 2017, \actaa, 67,
  363

\bibitem[{{Soszy{\'n}ski} {et~al.}(2019){Soszy{\'n}ski}, {Udalski},
  {Szyma{\'n}ski}, {Pietrukowicz}, {Skowron}, {Skowron}, {Poleski},
  {Koz{\l}owski}, {Mr{\'o}z}, {Ulaczyk}, {Rybicki}, {Iwanek}, \&
  {Wrona}}]{Soszynski2019AcA....69...87S}
{Soszy{\'n}ski}, I., {Udalski}, A., {Szyma{\'n}ski}, M.~K., {et~al.} 2019,
  \actaa, 69, 87

\bibitem[{Tonry {et~al.}(2018)Tonry, Denneau, Flewelling, Heinze, Onken,
  Smartt, Stalder, Weiland, \& Wolf}]{Tonry:2018}
Tonry, J.~L., Denneau, L., Flewelling, H., {et~al.} 2018, The Astrophysical
  Journal, 867, 105.
\newblock \url{https://doi.org/10.3847/1538-4357/aae386}

\bibitem[{{Torrealba} {et~al.}(2016{\natexlab{a}}){Torrealba}, {Koposov},
  {Belokurov}, \& {Irwin}}]{Torrealba:2016a}
{Torrealba}, G., {Koposov}, S.~E., {Belokurov}, V., \& {Irwin}, M.
  2016{\natexlab{a}}, \mnras, 459, 2370

\bibitem[{{Torrealba} {et~al.}(2016{\natexlab{b}}){Torrealba}, {Koposov},
  {Belokurov}, {Irwin}, {Collins}, {Spencer}, {Ibata}, {Mateo}, {Bonaca}, \&
  {Jethwa}}]{Torrealba:2016b}
{Torrealba}, G., {Koposov}, S.~E., {Belokurov}, V., {et~al.}
  2016{\natexlab{b}}, \mnras, 463, 712

\bibitem[{{Torrealba} {et~al.}(2018){Torrealba}, {Belokurov}, {Koposov},
  {Bechtol}, {Drlica-Wagner}, {Olsen}, {Vivas}, {Yanny}, {Jethwa}, {Walker},
  {Li}, {Allam}, {Conn}, {Gallart}, {Gruendl}, {James}, {Johnson}, {Kuehn},
  {Kuropatkin}, {Martin}, {Martinez-Delgado}, {Nidever}, {No{\"e}l}, {Simon},
  {Stringfellow}, \& {Tucker}}]{Torrealba:2018a}
{Torrealba}, G., {Belokurov}, V., {Koposov}, S.~E., {et~al.} 2018, \mnras, 475,
  5085

\bibitem[{{Torrealba} {et~al.}(2019){Torrealba}, {Belokurov}, {Koposov}, {Li},
  {Walker}, {Sanders}, {Geringer-Sameth}, {Zucker}, {Kuehn}, \&
  {Evans}}]{Torrealba:2019}
---. 2019, \mnras, 1548

\bibitem[{{van der Marel} {et~al.}(2002){van der Marel}, {Alves}, {Hardy}, \&
  {Suntzeff}}]{lmcrv}
{van der Marel}, R.~P., {Alves}, D.~R., {Hardy}, E., \& {Suntzeff}, N.~B. 2002,
  \aj, 124, 2639

\bibitem[{van~der Marel \& Kallivayalil(2014)}]{van_der_Marel_2014}
van~der Marel, R.~P., \& Kallivayalil, N. 2014, The Astrophysical Journal, 781,
  121.
\newblock \url{https://doi.org/10.1088\%2F0004-637x\%2F781%2F2\%2F121}

\bibitem[{{Van Der Walt} {et~al.}(2011){Van Der Walt}, {Colbert}, \&
  {Varoquaux}}]{numpy:2011}
{Van Der Walt}, S., {Colbert}, S.~C., \& {Varoquaux}, G. 2011, Computing in
  Science \& Engineering, 13, 22

\bibitem[{{Wan} {et~al.}(2020){Wan}, {Guglielmo}, {Lewis}, {Mackey}, \&
  {Ibata}}]{Skymapper:2020}
{Wan}, Z., {Guglielmo}, M., {Lewis}, G.~F., {Mackey}, D., \& {Ibata}, R.~A.
  2020, \mnras, 492, 782

\bibitem[{Webb {et~al.}(2014)Webb, Leigh, Sills, Harris, \&
  Hurley}]{10.1093/mnras/stu961}
Webb, J.~J., Leigh, N., Sills, A., Harris, W.~E., \& Hurley, J.~R. 2014,
  Monthly Notices of the Royal Astronomical Society, 442, 1569.
\newblock \url{https://doi.org/10.1093/mnras/stu961}

\bibitem[{{Weisz} {et~al.}(2013){Weisz}, {Dolphin}, {Skillman}, {Holtzman},
  {Dalcanton}, {Cole}, \& {Neary}}]{LMCSFH_1}
{Weisz}, D.~R., {Dolphin}, A.~E., {Skillman}, E.~D., {et~al.} 2013, \mnras,
  431, 364

\bibitem[{{Wenger} {et~al.}(2000){Wenger}, {Ochsenbein}, {Egret}, {Dubois},
  {Bonnarel}, {Borde}, {Genova}, {Jasniewicz}, {Lalo{\"e}}, {Lesteven}, \&
  {Monier}}]{SIMBAD}
{Wenger}, M., {Ochsenbein}, F., {Egret}, D., {et~al.} 2000, \aaps, 143, 9

\bibitem[{{Willman} {et~al.}(2011){Willman}, {Geha}, {Strader}, {Strigari},
  {Simon}, {Kirby}, {Ho}, \& {Warres}}]{2011AJ....142..128W}
{Willman}, B., {Geha}, M., {Strader}, J., {et~al.} 2011, \aj, 142, 128

\bibitem[{Willman \& Strader(2012)}]{Willman:2012}
Willman, B., \& Strader, J. 2012, The Astronomical Journal, 144, 76.
\newblock \url{https://doi.org/10.1088%2F0004-6256%2F144%2F3%2F76}

\bibitem[{{Willman} {et~al.}(2005{\natexlab{a}}){Willman}, {Blanton}, {West},
  {Dalcanton}, {Hogg}, {Schneider}, {Wherry}, {Yanny}, \&
  {Brinkmann}}]{Willman2005AJ....129.2692W}
{Willman}, B., {Blanton}, M.~R., {West}, A.~A., {et~al.} 2005{\natexlab{a}},
  \aj, 129, 2692

\bibitem[{{Willman} {et~al.}(2005{\natexlab{b}}){Willman}, {Dalcanton},
  {Martinez-Delgado}, {West}, {Blanton}, {Hogg}, {Barentine}, {Brewington},
  {Harvanek}, {Kleinman}, {Krzesinski}, {Long}, {Neilsen}, {Nitta}, \&
  {Snedden}}]{Willman:2005}
{Willman}, B., {Dalcanton}, J.~J., {Martinez-Delgado}, D., {et~al.}
  2005{\natexlab{b}}, \apjl, 626, L85

\bibitem[{Wolf {et~al.}(2018)Wolf, Onken, Luvaul, Schmidt, Bessell, Chang,
  Da~Costa, Mackey, Martin-Jones, Murphy, \& et~al.}]{Wolf_2018}
Wolf, C., Onken, C.~A., Luvaul, L.~C., {et~al.} 2018, Publications of the
  Astronomical Society of Australia, 35, doi:10.1017/pasa.2018.5.
\newblock \url{http://dx.doi.org/10.1017/pasa.2018.5}

\bibitem[{{York} {et~al.}(2000){York}, {Adelman}, {Anderson}, {Anderson},
  {Annis}, {Bahcall}, {Bakken}, {Barkhouser}, {Bastian}, {Berman}, {Boroski},
  {Bracker}, {Briegel}, {Briggs}, {Brinkmann}, {Brunner}, {Burles}, {Carey},
  {Carr}, {Castander}, {Chen}, {Colestock}, {Connolly}, {Crocker}, {Csabai},
  {Czarapata}, {Davis}, {Doi}, {Dombeck}, {Eisenstein}, {Ellman}, {Elms},
  {Evans}, {Fan}, {Federwitz}, {Fiscelli}, {Friedman}, {Frieman}, {Fukugita},
  {Gillespie}, {Gunn}, {Gurbani}, {de Haas}, {Haldeman}, {Harris}, {Hayes},
  {Heckman}, {Hennessy}, {Hindsley}, {Holm}, {Holmgren}, {Huang}, {Hull},
  {Husby}, {Ichikawa}, {Ichikawa}, {Ivezi{\'c}}, {Kent}, {Kim}, {Kinney},
  {Klaene}, {Kleinman}, {Kleinman}, {Knapp}, {Korienek}, {Kron}, {Kunszt},
  {Lamb}, {Lee}, {Leger}, {Limmongkol}, {Lindenmeyer}, {Long}, {Loomis},
  {Loveday}, {Lucinio}, {Lupton}, {MacKinnon}, {Mannery}, {Mantsch}, {Margon},
  {McGehee}, {McKay}, {Meiksin}, {Merelli}, {Monet}, {Munn}, {Narayanan},
  {Nash}, {Neilsen}, {Neswold}, {Newberg}, {Nichol}, {Nicinski}, {Nonino},
  {Okada}, {Okamura}, {Ostriker}, {Owen}, {Pauls}, {Peoples}, {Peterson},
  {Petravick}, {Pier}, {Pope}, {Pordes}, {Prosapio}, {Rechenmacher}, {Quinn},
  {Richards}, {Richmond}, {Rivetta}, {Rockosi}, {Ruthmansdorfer}, {Sandford},
  {Schlegel}, {Schneider}, {Sekiguchi}, {Sergey}, {Shimasaku}, {Siegmund},
  {Smee}, {Smith}, {Snedden}, {Stone}, {Stoughton}, {Strauss}, {Stubbs},
  {SubbaRao}, {Szalay}, {Szapudi}, {Szokoly}, {Thakar}, {Tremonti}, {Tucker},
  {Uomoto}, {Vanden Berk}, {Vogeley}, {Waddell}, {Wang}, {Watanabe},
  {Weinberg}, {Yanny}, {Yasuda}, \& {SDSS Collaboration}}]{York:2000}
{York}, D.~G., {Adelman}, J., {Anderson}, Jr., J.~E., {et~al.} 2000, \aj, 120,
  1579

\bibitem[{{Zucker} {et~al.}(2006{\natexlab{a}}){Zucker}, {Belokurov}, {Evans},
  {Kleyna}, {Irwin}, {Wilkinson}, {Fellhauer}, {Bramich}, {Gilmore}, {Newberg},
  {Yanny}, {Smith}, {Hewett}, {Bell}, {Rix}, {Gnedin}, {Vidrih}, {Wyse},
  {Willman}, {Grebel}, {Schneider}, {Beers}, {Kniazev}, {Barentine},
  {Brewington}, {Brinkmann}, {Harvanek}, {Kleinman}, {Krzesinski}, {Long},
  {Nitta}, \& {Snedden}}]{ZuckerApJ...650L..41Z}
{Zucker}, D.~B., {Belokurov}, V., {Evans}, N.~W., {et~al.} 2006{\natexlab{a}},
  \apjl, 650, L41

\bibitem[{{Zucker} {et~al.}(2006{\natexlab{b}}){Zucker}, {Belokurov}, {Evans},
  {Wilkinson}, {Irwin}, {Sivarani}, {Hodgkin}, {Bramich}, {Irwin}, {Gilmore},
  {Willman}, {Vidrih}, {Fellhauer}, {Hewett}, {Beers}, {Bell}, {Grebel},
  {Schneider}, {Newberg}, {Wyse}, {Rockosi}, {Yanny}, {Lupton}, {Smith},
  {Barentine}, {Brewington}, {Brinkmann}, {Harvanek}, {Kleinman}, {Krzesinski},
  {Long}, {Nitta}, \& {Snedden}}]{Zucker2006ApJ...643L.103Z}
---. 2006{\natexlab{b}}, \apjl, 643, L103

\end{thebibliography}
